\documentclass[twocolumn]{aastex701}
\usepackage{makecell}
\usepackage{xcolor}
\usepackage{enumitem}
\usepackage[T1]{fontenc}

\begin{document}

\title{X-ray and optical decline of the intermediate mass black hole HLX-1}

\author[orcid=0000-0002-4622-796X,sname='Soria']{Roberto Soria}
\affiliation{INAF-Osservatorio Astrofisico di Torino, Strada Osservatorio 20, I-10025 Pino Torinese, Italy}
\affiliation{College of Astronomy and Space Sciences, University of the Chinese Academy of Sciences, Beijing 100049, People's Republic of China}
\affiliation{Sydney Institute for Astronomy, School of Physics A28, The, University of Sydney, Sydney, NSW, Australia}
\email[show]{roberto.soria@inaf.it}  

\author[orcid=0009-0006-8094-1362, sname='Garbaccio Gili']{Maya Garbaccio Gili} 
\affiliation{INAF-Istituto di Astrofisica e Planetologia Spaziali, Via del Fosso del Cavaliere 100, I-00133 Roma, Italy}
\affiliation{Dipartimento di Fisica, Universit\'a degli Studi di Torino, Via Pietro Giuria 1, I-10125 Torino, Italy}
\affiliation{Dipartimento di Fisica, Universit\'a di Roma Tor Vergata, Via della Ricerca Scientifica 1, I-00133 Roma, Italy}
\email[show]{maya.garbacciogili@inaf.it}

\author[orcid=0000-0002-1704-9850,sname=Massaro]{Francesco Massaro}
\affiliation{Dipartimento di Fisica, Universit\'a degli Studi di Torino, Via Pietro Giuria 1, I-10125 Torino, Italy}
\affiliation{Istituto Nazionale di Fisica Nucleare, Sezione di Torino, Via Pietro Giuria 1, I-10125 Torino, Italy}
\affiliation{INAF-Osservatorio Astrofisico di Torino, Strada Osservatorio 20, I-10025 Pino Torinese, Italy}
\email{f.massaro@unito.it}

\author[orcid=0000-0002-6496-9414,sname=Graham]{Alister W. Graham}
\affiliation{Centre for Astrophysics and Supercomputing, Swinburne University of Technology, Hawthorn, VIC, Australia}
\email{agraham@astro.swin.edu.au}

\author[orcid=0000-0002-6516-1329,sname=Zampieri]{Luca Zampieri}
\affiliation{INAF-Osservatorio Astronomico di Padova, Vicolo dell'Osservatorio 5, I-35122, Padova, Italy}
\email{luca.zampieri@inaf.it}


\begin{abstract}

HLX-1 is a prominent intermediate-mass black hole (IMBH) candidate, historically exhibiting recurrent X-ray outbursts with spectral state transitions analogous to those observed in stellar-mass black holes. Here, we present new \textit{Hubble Space Telescope}, \textit{Chandra}, and \textit{Swift} observations from 2018--2022 to characterise the late-time flux decline. HLX-1 has persisted in a low X-ray luminosity state ($L_{\rm X} \approx $ a few $ \times 10^{39}$ erg s$^{-1}$) since the end of its last outburst in 2017. We observe a significant decoupling between the X-ray and optical/UV emission: while the X-rays have faded by at least two orders of magnitude from peak outburst luminosity (in 2010) to the current low state, the optical/UV flux has declined much more slowly over the same time. This results in an X-ray/optical luminosity ratio inconsistent with X-ray reprocessing in a standard accretion disk, as this would require an unphysical reprocessing fraction $>$100$\%$ at late times. Instead, we find that the optical/UV evolution is well-fit by a cooling, expanding photosphere ($T \approx 30,000$ K), similar to the late-stage evolution seen in tidal disruption events (TDEs). The redder component of the optical emission is instead consistent with the old stellar population of a massive star cluster (IMBH host). The pre-2017 X-ray bursting phase is consistent with simulations of disk instabilities in TDE evolution: this strengthens the scenario of HLX-1 as an IMBH TDE. Furthermore, our observations resolve the morphology and flux of the mysterious far-UV emitter, seen in projection next to HLX-1, into a ring-like star-forming structure. We re-assess the possibility that HLX-1 and its host star cluster are physically associated with this starburst dwarf, perhaps via a high-speed collision. 

\end{abstract}



\section{Introduction}

The soft X-ray transient 2XMM J011028.1$-$460421 in the galaxy cluster Abell 2877 \citep{1992AJ....104..495M} is one of the best studied and most convincing intermediate-mass black hole (IMBH) candidates. Originally discovered from the {\it XMM-Newton} serendipitous source catalog  \citep{farrell09}, it was then extensively studied with all major X-ray facilities, especially the {\it{Chandra X-Ray Observatory}}, {\it{XMM-Newton}}, and the {\it{Neil Gehrels Swift Observatory}}. It is located at RA(J2000) $= 01^{h} 10^m 28^s.28$, Dec(J2000) $= -46^\circ 04^\prime 22$\farcs3, and is seen projected $\approx$8$^{\prime\prime}$ ($\approx 3.6$ kpc) north-east of the nucleus of the lenticular galaxy ESO\,243-49.  
This galaxy has a heliocentric recession speed of $\approx$ 6700 km s$^{-1}$ \citep{caldwell97,webb17}, corresponding to a 
redshift $z = 0.0224 \pm 0.0001$\footnote{NASA/IPAC Extragalactic Database, \url{https://ned.ipac.caltech.edu}.}, a luminosity distance $d_{\rm L} \approx 98$ Mpc and an angular diameter distance $d_{\rm A} \approx 93$ Mpc. The status of IMBH candidate for 2XMM J011028.1$-$460421 rests on the hypothesis that the X-ray source is physically located at a similar distance as ESO\,243-49 and is not, instead, a foreground Galactic X-ray binary or a background AGN. This hypothesis is supported by the detection of an emission line at $\lambda \approx 6720$\AA\ \citep{wiersema10,soria13}, interpreted as H$\alpha$ redshifted by $z = 0.0239$, from its optical counterpart. Based on this {\it approximate} coincidence of the two redshifts, the X-ray source is better known in the literature as ESO\,243-49 HLX-1 (hereinafter, HLX-1). 

HLX-1 was observed for the first time in archival {\it{XMM-Newton}} observations taken on 2004 November 23 (MJD 53332), showing a soft X-ray spectrum and an unabsorbed 0.3--10 keV luminosity $L_{\rm X} \approx 10^{42}$ erg s$^{-1}$ \citep{farrell09}. Seen again in a high/soft state in 2008 November, 
it has been monitored at regular intervals since then. 
It went into outburst eight times between 2008 and 2017 \citep{farrell09,webb17,soria17}. Each outburst peaked at a roughly similar $L_{\rm X} \approx 10^{42}$ erg s$^{-1}$, although the outburst duration and fluence decreased over the outburst sequence, and the time intervals between outbursts increased. At the peak of each outburst, the X-ray spectrum was soft and dominated by a thermal component, interpreted as emission from a standard accretion disk \citep{servillat11,davis11,farrell12,godet12,soria17}. This behaviour is similar to the high/soft state in stellar-mass black hole (BH) outbursts. In between outbursts, the spectrum was harder, consistent with a power-law and at least phenomenologically similar to the low/hard state of stellar-mass BHs \citep{servillat11,yan15}, even though the physics may be different \citep{lasota11}. The detection of radio flares at the transition from the hard to the soft state \citep{webb12} is similar to what happens in stellar-mass BHs. The peak colour temperature $kT_{\rm in} \approx 0.25$ keV in the high/soft state, together with the peak luminosity of $\approx$10$^{42}$ erg s$^{-1}$, are consistent with a BH of mass $M \approx 2 \times 10^4 M_\odot$ \citep{godet12,soria17}. Beyond this general understanding, many things remain unclear in this system: {\it{e.g.}}, what triggered the recurrent outbursts (disk instabilities or partial tidal disruption of an orbiting star?); why the oscillation has now stopped; why the outburst and recurrence time scales were similar to those seen in stellar-mass BHs rather than those expected for a much larger IMBH accretion disk; and the age and mass of the host stellar environment.  

HLX-1 has a point-like (size $\lesssim$ 0\farcs1 in {\it{Hubble Space Telescope}} [{\it{HST}}] images) optical counterpart, whose nature also remains unclear \citep{soria10,farrell12,farrell14,soria17}. Its dimming and reddening over the years suggests that the dominant component, at least in the earliest {\it{HST}} observations, is hot gas associated with the X-ray outburst (for example an irradiated accretion disk), rather than stellar emission from a host star cluster \citep{soria17}. The candidate H$\alpha$ emission line mentioned earlier, on which the distance estimate is based, was seen only in the 2009 and 2012 Very Large Telescope (VLT) FORS2 observations; it had disappeared in the VLT/X-shooter and 2014 VLT/MUSE spectra \citep{webb17}.


Here, we investigate the X-ray and optical decline of HLX-1 after its last recorded outburst. We will:
\begin{itemize}[noitemsep, topsep=0pt, parsep=0pt, partopsep=0pt, leftmargin=8pt]
\item disentangle and compare the X-ray luminosity of HLX-1 in the low/hard state with the extended, unresolved emission from the X-ray binary population of ESO\,243-49;
\item assess the nature of the optical emission, testing whether it is consistent with a decreasing bluer component from an irradiated accretion disk, plus a constant redder component from a star cluster;
\item re-consider possible connections between HLX-1 and a ring-like, star-forming complex, seen in projection between HLX-1 and the nucleus of ESO\,243-49, but apparently located well behind them \citep{wiersema10,soria13,webb17}.
\end{itemize}

We adopt cgs units for numerical results and we assume a
flat cosmology with $H_0 = 69.6$ km s$^{-1}$ Mpc$^{-1}$, $\Omega_{\mathrm{M}} = 0.286$ and $\Omega_\Lambda = 0.714$ \citep{bennett14}.

\begin{deluxetable*}{lccccccc}
\digitalasset
\tablewidth{0pt}
\tablecaption{{\it{HST}} observation log and observed brightness of HLX-1. 
}
\label{tab:hst_mag_results}
\tablehead{
\colhead{\makecell[c]{Date \\ ~ }} & \colhead{\makecell[c]{MJD$_{\mathrm{mid}}$ \\ ~ }}
& \colhead{\makecell[c]{$\Delta t$ after peak$^a$ \\ (d)}} & 
\colhead{\makecell[c]{Instrument \\ ~ }} &
\colhead{\makecell[c]{Filter \\ ~ }} & \colhead{ \makecell[c]{Pivot wavelength$^b$ \\ (\AA)}} & \colhead{ \makecell[c]{Brightness \\ (Vegamag)} } & \colhead{ \makecell[c]{Brightness \\ (ABmag)} }
}
\startdata\\[-5pt]
\makecell[c]{2010-09-13 \\ 2010-09-23 \\ \\ \\ \\ ~} & 
\makecell[c]{55452.28 \\ 55462.60 \\ 55462.65 \\ 55462.66 \\ 55462.67 \\ 55462.72} &
\makecell[c]{12 \\ 22 \\ \\ \\ \\ ~} & 
\makecell[c]{ACS-SBC \\ WFC3-UVIS \\ \\ \\ \\ WFC3-IR} &
\makecell[c]{F140LP \\ F300X \\ F390W \\ F555W \\ F775W \\ F160W} & \makecell[c]{1519.3 \\ 2820.5 \\ 3923.7 \\ 5308.4 \\ 7651.4 \\ 15369.2} & \makecell[c]{22.20 $\pm$ 0.08 \\ 22.61 $\pm$ 0.02 \\ 23.88 $\pm$ 0.06 \\ 24.04 $\pm$ 0.14 \\ 23.76 $\pm$ 0.04 \\ 23.04 $\pm$ 0.47} & 
\makecell[c]{24.34 $\pm$ 0.08 \\ 24.01 $\pm$ 0.02 \\ 24.09 $\pm$ 0.06 \\ 24.01 $\pm$ 0.14 \\ 24.15 $\pm$ 0.04 \\ 24.31 $\pm$ 0.47} \\  
\tableline\\[-5pt]
\makecell[c]{2012-11-19 \\ \\ \\ \\ \\ \\ \\ \\ ~} & 
\makecell[c]{56250.66 \\ 56250.76 \\ 56250.77 \\ 56250.89 \\ 56250.91 \\ 56250.96 \\ 56250.97 \\ 56250.82 \\ 56250.84} &
\makecell[c]{78 \\ \\ \\ \\ \\ \\ \\ \\ ~} & 
\makecell[c]{ACS-SBC \\ WFC3-UVIS \\ \\ \\ \\ \\ \\ WFC3-IR \\ ~} &
\makecell[c]{F140LP \\ F300X \\ F336W \\ F390W \\ F555W \\ F621M \\ F775W \\ F105W \\ F160W} & 
\makecell[c]{1519.3 \\ 2820.5 \\ 3354.5 \\ 3923.7 \\ 5308.4 \\ 6218.9 \\ 7651.4 \\ 10551.1 \\ 15369.2} & 
\makecell[c]{22.61 $\pm$ 0.08 \\ 23.23 $\pm$ 0.02 \\ 23.53 $\pm$ 0.04 \\ 23.37 $\pm$ 0.06 \\ 24.42 $\pm$ 0.14 \\ 24.28 $\pm$ 0.22 \\ 24.00 $\pm$ 0.04 \\ 23.83 $\pm$ 0.17 \\ 23.26 $\pm$ 0.47 } & 
\makecell[c]{24.75 $\pm$ 0.08 \\ 24.63 $\pm$ 0.02 \\ 24.70 $\pm$ 0.04 \\ 24.57 $\pm$ 0.06 \\ 24.39 $\pm$ 0.14 \\ 24.43 $\pm$ 0.22 \\ 24.39 $\pm$ 0.04 \\ 24.50 $\pm$ 0.17 \\ 24.54 $\pm$ 0.47} \\
\tableline\\[-5pt]
\makecell[c]{2013-07-05 \\ \\ \\ \\ \\ \\ 2013-07-06 \\ \\ ~} & 
\makecell[c]{56478.14 \\ 56478.80 \\ 56478.82 \\ 56478.93 \\ 56478.95 \\ 56479.00 \\ 56479.01 \\ 56478.86 \\ 56478.88 } &
\makecell[c]{272 \\ \\ \\ \\ \\ \\ 273 \\ \\~ } & 
\makecell[c]{ACS-SBC \\ WFC3-UVIS \\ \\ \\ \\ \\ \\ WFC3-IR \\~} &
\makecell[c]{F140LP \\ F300X \\ F336W \\ F390W \\ F555W \\ F621M \\ F775W \\ F105W \\ F160W} & 
\makecell[c]{1528.0 \\ 2814.8 \\ 3354.5 \\ 3923.7 \\ 5308.4 \\ 6218.9 \\ 7651.4 \\ 10551.1 \\ 15369.2} & 
\makecell[c]{23.65 $\pm$ 0.08 \\ 24.12 $\pm$ 0.02 \\ 23.93 $\pm$ 0.04 \\ 25.38 $\pm$ 0.06 \\ 25.06 $\pm$ 0.14 \\ 24.58 $\pm$ 0.22 \\ 24.32 $\pm$ 0.04 \\ 24.09 $\pm$ 0.17 \\ 23.30 $\pm$ 0.47 } & 
\makecell[c]{25.79 $\pm$ 0.08\\ 25.52 $\pm$ 0.02 \\ 25.10 $\pm$ 0.04 \\ 25.58 $\pm$ 0.06 \\ 25.03 $\pm$ 0.14 \\ 24.73 $\pm$ 0.22 \\ 24.71 $\pm$ 0.04 \\ 24.76 $\pm$ 0.17 \\ 24.57 $\pm$ 0.47} \\
\tableline\\[-5pt]
\makecell[c]{2018-10-29 \\ 2018-10-30  \\ \\ 2018-10-31 \\ \\ \\ \\ \\~ } & 
\makecell[c]{58420.29 \\ 58421.92 \\ 58422.00 \\ 58422.03 \\ 58422.08 \\ 58422.12 \\ 58422.16 \\ 58422.21 \\ 58420.23 } &
\makecell[c]{552 \\ 553  \\ \\ 554 \\ \\ \\ \\ \\~ } &
\makecell[c]{WFC3-IR \\ ACS-SBC  \\ WFC3-UVIS\\ \\ \\ \\ \\ \\ WFC3-IR} &
\makecell[c]{F160W\\ F140LP \\ F300X \\ F336W \\ F390W \\ F555W \\ F621M \\ F775W \\ F105W } & 
\makecell[c]{15369.2 \\  1519.3 \\ 2820.5 \\ 3354.5 \\ 3923.7 \\ 5308.4 \\ 6218.9 \\ 7651.4 \\ 10551.1 } & 
\makecell[c]{23.46 $\pm$ 0.47 \\ 23.79 $\pm$ 0.08 \\ 24.38 $\pm$ 0.02 \\ 24.52 $\pm$ 0.04 \\ 25.42 $\pm$ 0.06 \\ 25.11 $\pm$ 0.14 \\ 24.69 $\pm$ 0.22 \\ 24.30 $\pm$ 0.04 \\ 23.89 $\pm$ 0.17 } & 
\makecell[c]{24.74 $\pm$ 0.47 \\ 25.93 $\pm$ 0.08\\ 25.79 $\pm$ 0.02 \\ 25.68 $\pm$ 0.04 \\ 25.62 $\pm$ 0.06 \\ 25.08 $\pm$ 0.14 \\ 24.84 $\pm$ 0.22 \\ 24.69 $\pm$ 0.04 \\ 24.55 $\pm$ 0.17 } \\
\tableline\\[-5pt]
\makecell[c]{2022-05-22 \\ \\ \\ \\ \\ \\~ } & 
\makecell[c]{59721.58 \\ 59721.64 \\ 59721.68 \\ 59721.71 \\ 59721.77 \\ 59721.81 \\ 59721.85} & 
\makecell[c]{1853 \\ \\ \\ \\ \\ \\~ } & 
\makecell[c]{ACS-SBC \\ WFC3-UVIS\\ \\ \\ \\ \\~ } & 
\makecell[c]{F140LP \\ F300X \\ F336W \\ F390W \\ F555W \\ F621M \\ F775W} & 
\makecell[c]{1519.3 \\ 2820.5 \\ 3354.5 \\ 3923.7 \\ 5308.4 \\ 6218.9 \\ 7651.4} & 
\makecell[c]{23.85 $\pm$ 0.08 \\ 24.60 $\pm$ 0.02 \\ 24.55 $\pm$ 0.04 \\ 25.58 $\pm$ 0.06 \\ 25.24 $\pm$ 0.14 \\ 24.95 $\pm$ 0.22 \\ 24.39 $\pm$ 0.04 } & 
\makecell[c]{26.00 $\pm$ 0.08 \\ 26.00 $\pm$ 0.02 \\ 25.71 $\pm$ 0.04 \\ 25.77 $\pm$ 0.06 \\ 25.21 $\pm$ 0.14 \\ 25.10 $\pm$ 0.22 \\ 24.78 $\pm$ 0.04 }
\enddata
\tablecomments{$^a$: time between the peak of the preceding X-ray outburst and the {\it{HST}} observation.\\ $^b$: the ``pivot wavelength'' ($\lambda_{\mathrm{p}}$) is a source-independent measure of the characteristic wavelength of a bandpass, and is defined in \cite{tokunaga05}. For the ACS filter F140LP, $\lambda_{\mathrm{p}}$ is taken directly from the headers of the image files. For all the WFC3 filters, $\lambda_{\mathrm{p}}$ are the UVIS1 values (vacuum wavelengths) from \cite{calamida22}; see also Chapter 6.5 in the WFC3 Instrument Handbook, \url{https://hst-docs.stsci.edu/wfc3ihb}  }
\end{deluxetable*}

\section{Observations and data analysis}

\subsection{X-ray data: Chandra and Swift}

The field of HLX-1 was imaged with the back-illuminated S3 chip of the Advanced CCD Imaging Spectrometer (ACIS) on board {\it{Chandra}} on 2010 September 6 (for 10\,ks), on 2020 September 16 (for 20\,ks) and on 2022 May 23,24,25 (for 37\,ks, 20\,ks, and 20\,ks, respectively).
We downloaded the data from the Chandra Data Archive. 
For all data analysis, we used standard tools from the Chandra Interactive Analysis of Observations ({\sc{ciao}}) package \citep{fruscione06} version 4.17.0, with the latest version of the Calibration Database (CALDB version 4.12.0). 
Only the 2010 and 2022 observations provide useful data, with HLX-1 located in the back-illuminated ACIS-S3 chip. Instead, in 2020, HLX-1 was located in the S4 chip\footnote{The target of that observation was a different, unrelated sourse, 2MASX J01110461$-$4558435.}, 8\farcm4 away from the aimpoint; combined with the low state of the source and the relatively short exposure time, the result is that there is no significant detection for that epoch. Thus, we will only consider the 2010 and 2022 datasets.

We reprocessed the raw files with the {\it{chandra\_repro}} task, to create updated level-2 event files. We checked the data for possible background flares (none were found). We extracted images, light-curves and spectra (with associated responses and ancillary response files) with the standard tasks {\it{dmcopy}}, {\it{dmextract}} and {\it{specextract}}. In particular, for spectral extractions, we used a circular source region of radius 2\farcs5, and local background from an annulus between 3$^{\prime\prime}$ and 6$^{\prime\prime}$. The spectra from the 2010 observations
were re-grouped to a minimum of 15 counts per bin with the {\sc{ftools}} task {\it{grppha}} \citep{blackburn95,nasa14}, and fitted with the $\chi^2$ statistics. The spectra from the 2022 observations do not have enough counts for $\chi^2$ fitting: thus, they were re-grouped to a minimum of 1 count per bin, and fitted with the Cash statistics \citep{cash79}. For spectral modelling, we used the {\sc{xspec}} fitting package \citep{arnaud96} Version 12.14.1. All spectral modelling was done on the 0.3--7 keV energy range, although fluxes and luminosities were then extrapolated to the standard 0.3--10 keV range.

The X-Ray Telescope (XRT) on board \textit{Swift} has observed the field of HLX-1 for the first time in 2008 October--November, and afterwards it has monitored it regularly since 2009 August, with about 800 observations until the end of 2025: an average cadence of almost one per week, and a total exposure time of 1.33 Ms. 
We used the online \textit{Swift}/XRT data products generator\footnote{{\url{https://www.swift.ac.uk/user_objects/}}} \citep{evans07,evans09} to extract a 0.3--10 keV light-curve. The light-curve was 
rebinned to a minimum of 15 counts per bin in the source region and a minimum detection significance 
of 2.5$\sigma$. 

The lower spatial resolution of \textit{Swift}/XRT (18$^{\prime\prime}$ half-power diameter) implies that it is impossible to resolve the extended emission from the central region of ESO\,243-49 from the point-like emission from HLX-1. The two contributions can instead be easily resolved by {\it{Chandra}}/ACIS. Since the galaxy contribution is expected to be constant, we used the flux/count-rate conversion tool Web{\sc{pimms}} \citep{mukai93} Version 4.15 to estimate the corresponding count rate of this component in \textit{Swift}/XRT. Then, the net contribution of HLX-1 is derived as the measured flux from \textit{Swift}/XRT minus the estimated contribution of the galaxy.


\begin{deluxetable}{c c}
\label{tab:source_radii}
\digitalasset
\tablewidth{0pt}
\tablecaption{Source extraction radii adopted for each {\it{HST}} filter.} 
\tablehead{
\colhead{\ \ \ \ \ \ \ \ Source Radius\ \ \ \ \ \ \ \ \ } & \colhead{Filters} 
}
\startdata
0\farcs16 & F555W, F621M, F775W \\
0\farcs20 & F300X, F336W, F390W \\
0\farcs26 & F105W, F160W \\
0\farcs40 & F140LP \\
\enddata
\tablecomments{The encircled energy fractions of the WFC3 UVIS and IR filters are listed at \url{https://www.stsci.edu/hst/instrumentation/wfc3/data-analysis/photometric-calibration/}. Those for the ACS SBC filter are at \url{https://www.stsci.edu/hst/instrumentation/acs/data-analysis/aperture-corrections}.}
\end{deluxetable}

\subsection{Optical data: HST}

HLX-1 was observed by $\textit{HST}$ on five epochs (Table 1): in 2010, 2012, 2013, 2018 and 2022. Each set of observations typically includes one far-UV image (F140LP filter) taken with the Advanced Camera for Surveys in the Solar Blind Channel (ACS-SBC), a set of optical images with the Wide Field Camera 3 (WFC3) in the Ultraviolet and Visible (UVIS) channel, and one or two images with the WFC3 InfraRed (IR) channel (except in 2022 when no WFC3-IR images were taken). An analysis of the long-term optical decline and colour evolution based on the 2010, 2012 and 2013 datasets was presented in \cite{soria17}. Here, we develop and extend the analysis, including the 2018 and 2022 datasets. 

We downloaded the data from the Mikulski Archive for Space Telescopes (MAST), and re-analysed all five datasets, for consistency. We used aperture photometry to measure the brightness of HLX-1, with SAOImage's {\sc{ds9}} package, version 8.6. To define the source extraction radii, we looked for a trade-off between maximising the statistical reliability and minimising the contamination from background emission; this corresponds to an encircled energy fraction of $\approx$80\%. Both the encircled energy distribution and the relative contributions of source and background emission vary from band to band; for instance, larger source extraction radii are acceptable in the near-UV bands where the diffuse background emission from the galaxy halo is less significant. We chose source extraction radii of 0\farcs16 for the optical UVIS filters, 0\farcs20 for the near-UV UVIS filters, 0\farcs26 for the IR filters, and 0\farcs40 for the far-UV ACS filter (Table \ref{tab:source_radii}). We also verified that our results are not significantly changed if we had used radii of 0\farcs16 or 0\farcs20 for all UVIS filters. 

The main source of measurement uncertainty comes from the estimate of the background level at the location of HLX-1. There is a strong diffuse background, particularly dominant in the redder bands, caused by the old stellar population of ESO\,243-49. The background emission is severely non-uniform, being stronger towards the center of the galaxy (south-west of HLX-1). To reduce this problem, we built smoothed images of the galaxy in each band, 
and determined the contour levels (isophotes) of the diffuse galaxy emission passing through the location of HLX-1 (neglecting the emission peak of HLX-1 itself). We then computed the background level from ellipsoidal regions along the same isophotes, centered about 1$^{\prime\prime}$ to the east and to the west of HLX-1, with a combined area at least three times the area of the source extraction circle. This is particularly important for the optical and near-IR filters. Instead, for the near-UV filters, the stellar halo emission of ESO\,243-49 is negligible, and the background subtraction can be done equally well with annuli around the source circle.

We divided the net count rates of HLX-1 measured in each observation by the appropriate encircled-energy fractions, to obtain infinite-aperture count rates. Finally, we converted the count rates to flux densities and apparent magnitudes in the Vega and AB systems, using the inverse sensitivities and zeropoints in the online tables\footnote{For WFC3: \url{https://www.stsci.edu/hst/instrumentation/wfc3/data-analysis/photometric-calibration/}; for ACS: \url{https://acszeropoints.stsci.edu/}.} on the Space Telescope Science Institute website. As a further check, essentially identical values of inverse sensitivities are also stored in the fits header of each image.  


\begin{figure}[t]
\hspace{-0.5cm}
    \includegraphics[width=0.36\textwidth, angle=270]{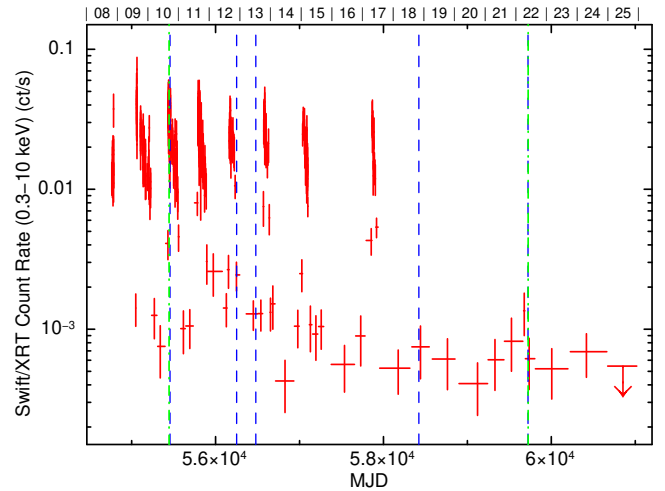}
\vspace{0.7cm}
    \caption{Red datapoints: {\it{Swift}}/XRT light-curve, rebinned to a minimum signal-to-noise ratio of 2.5. The centroid of the source extraction radius was fixed at the position of HLX-1; however, as a consequence of the large size of the XRT PSF, the X-ray flux is dominated by diffuse emission from ESO\,243-49 for count rates $\lesssim$10$^{-3}$ ct s$^{-1}$. Dashed blue lines mark the epochs of the five {\it{HST}} observations. Dot-dashed green lines mark the epochs of the 2010 and 2022 {\it{Chandra}} observations (approximately coincident with the first and last {\it{HST}} observations). Calendar years are labelled for convenience at the top of the plot.
    }
    \label{fig:xrt_lc}
\end{figure}

\section{X-ray Results}

\subsection{Light-curve from {\it{Swift}}/XRT monitoring}

{\it{Swift}}/XRT has monitored the evolution of HLX-1 weekly since 2008. After its last outburst in 2017, HLX-1 has remained in a low state. Individual snapshot XRT observations (typical exposure times between 1 and 3 ks) do not provide enough photons for significant detections after 2017. However, we can still build an X-ray light-curve by stacking a large number of observations: typically, about a year's worth of XRT exposures for a 3$\sigma$ detection. The resulting light-curve has been approximately constant since the end of the last outburst (Figure~\ref{fig:xrt_lc}), two orders of magnitude below the peak luminosity. However, {\it{Swift}}/XRT cannot resolve the contribution from HLX-1 and from the inner region of ESO\,243-49; therefore, we cannot conclude that HLX-1 has settled on a plateau. We used the {\it{Chandra}} data to achieve the following results: i) determine the luminosity and spectral properties of the unresolved galaxy emission; ii) compare the luminosity of HLX-1 and of the galaxy in the 2022 dataset; iii) estimate the contributions of the galaxy emission to the {\it{Swift}}/XRT count rate at the current level. The underlying assumption is that the galaxy contribution is not strongly variable from year to year.

\begin{figure*}[t]
    \centering
\hspace{-0.5cm}
    \includegraphics[width=0.33\textwidth]{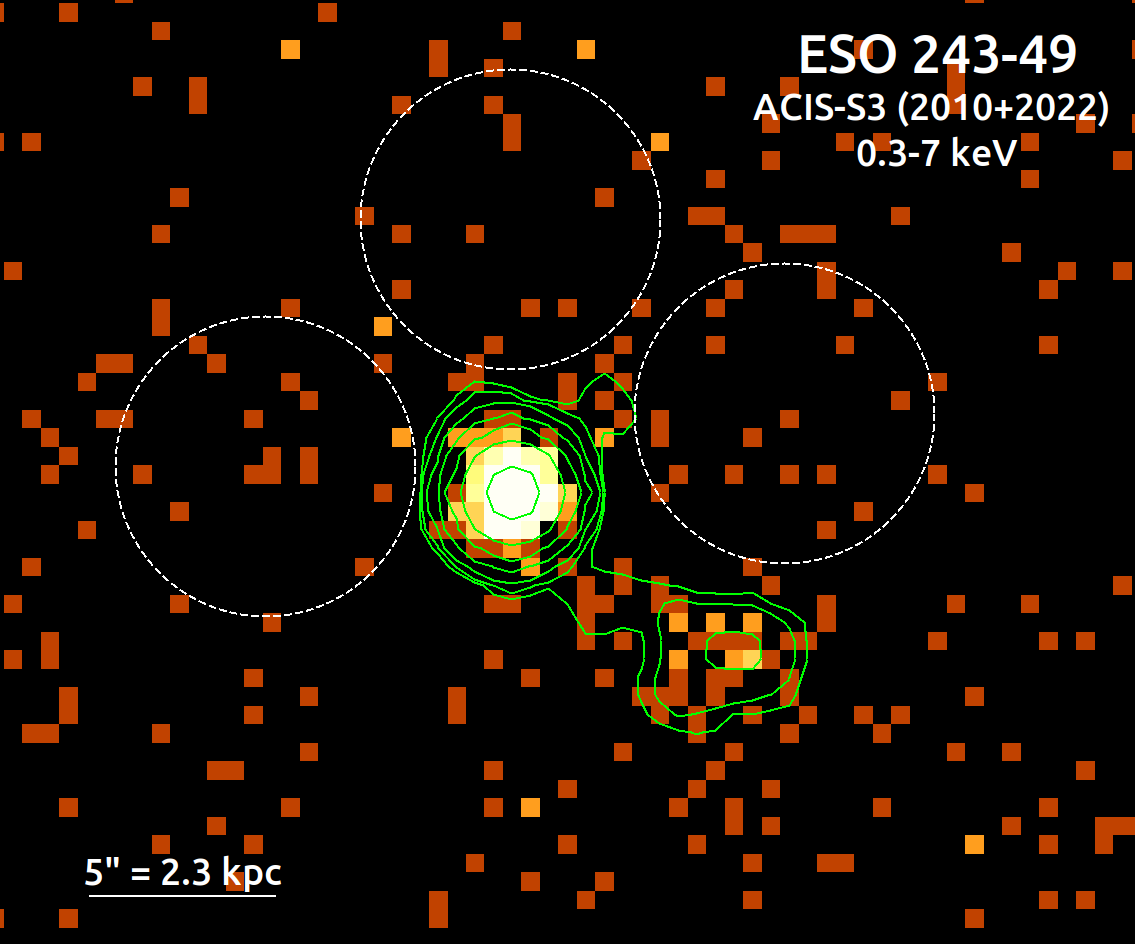}
    \includegraphics[width=0.33\textwidth]{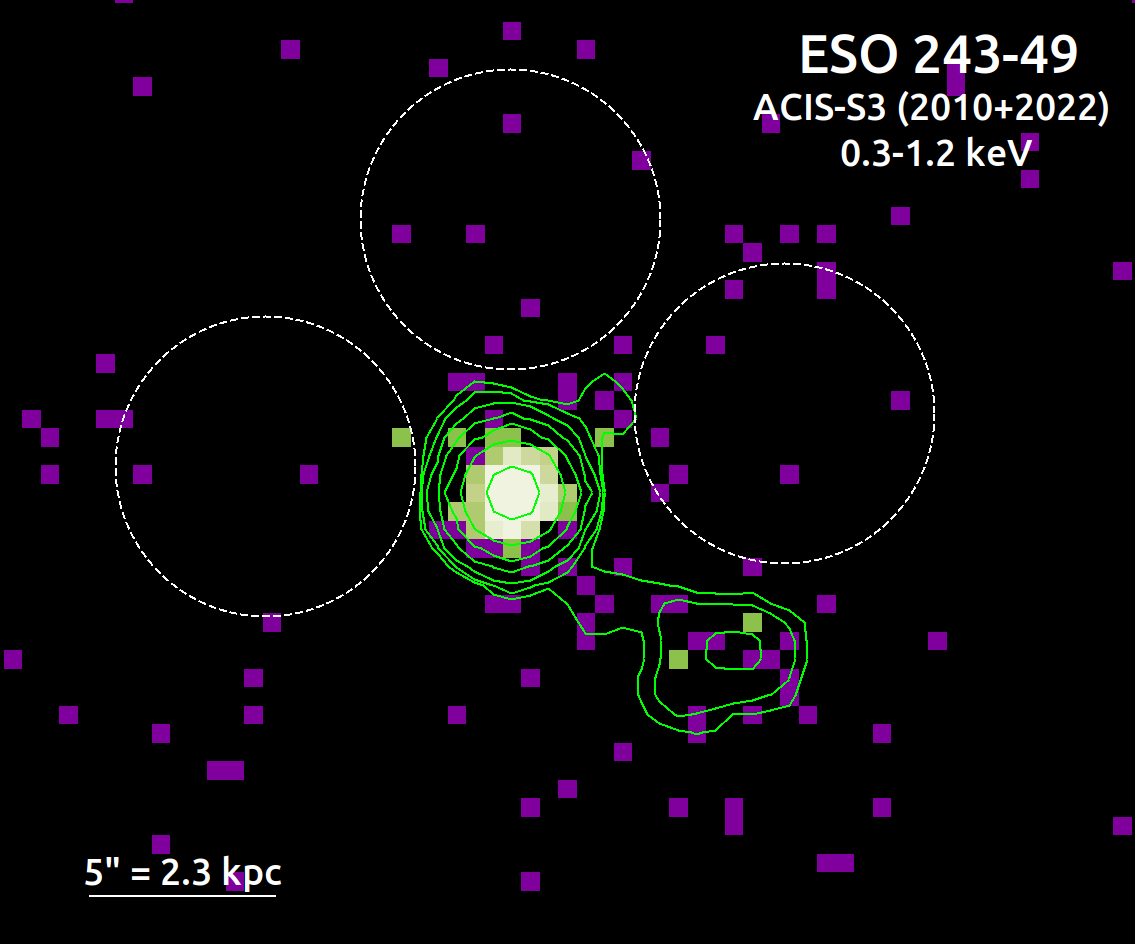}
    \includegraphics[width=0.33\textwidth]{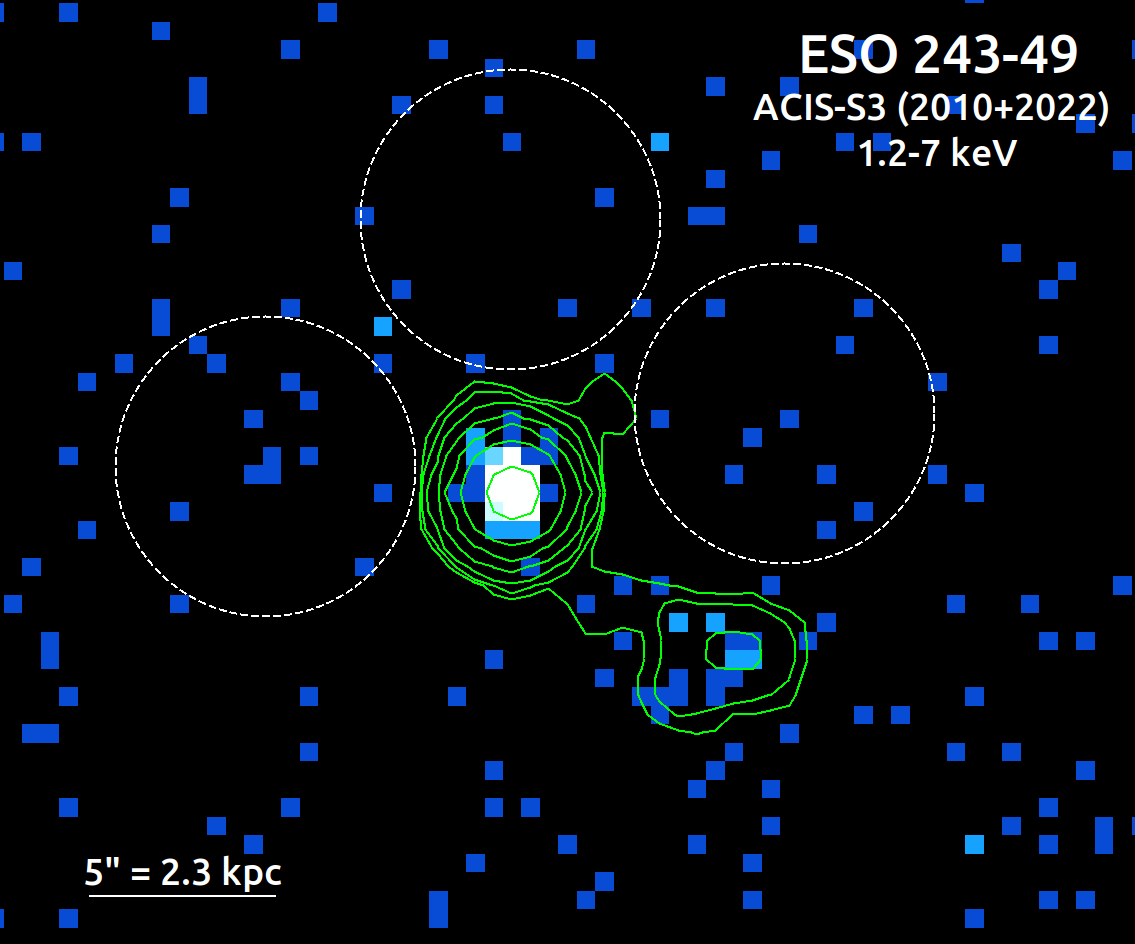}
    \caption{Left: {\it{Chandra}}/ACIS map in the 0.3--7.0 keV band, from the stacked 2010 plus 2022 data, with Gaussian-smoothed contour levels. HLX-1 is the upper (point-like) source. Dashed white circles represent the composite background region used for the spectral extraction of HLX-1. Middle: 0.3--1.2 keV map, from the same data; the contours are those for the full energy band. Right: 1.2--7.0 keV map, with full-band contours. In all panels, north is up and east to the left.
    }
    \label{fig:chandra_contours}
\end{figure*}

\begin{figure}[t]
    \centering
\hspace{-0.5cm}
    \includegraphics[width=0.48\textwidth]{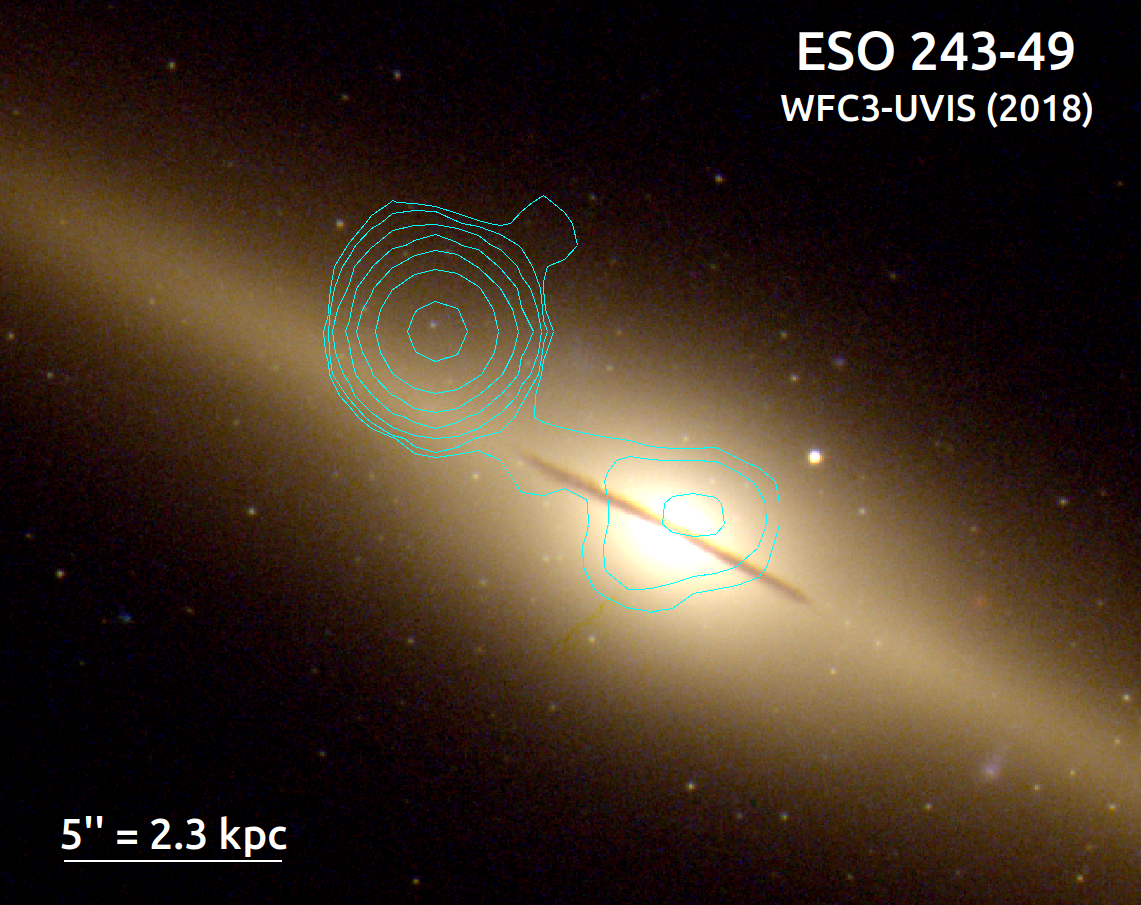}
    \caption{{\it{Chandra}}/ACIS count-rate contours (cyan) from the merged 2010 and 2022 datasets, overplotted on an {\it{HST}}/WFC3 image of ESO \,243-49 from 2018 (red = F775W; green = F555W; blue = F390W). The contours have been smoothed with a 3-pixel Gaussian kernel. While {\it{Chandra}} resolves the two X-ray contributions (point-like and strongly variable emission from HLX-1 on the left, extended and roughly constant from the galaxy bulge on the right), {\it{Swift}} can only monitor the sum of the two components. The faint star-like source near the peak of the {\it{Chandra}} contours is the optical counterpart of HLX-1.
    }
    \label{fig:chandra_hst}
\end{figure}

We used the {\sc{ciao}} task {\it{merge\_obs}} to create a stacked event file from the 2010 and 2022 {\it{Chandra}} observations. The merged image reveals two sources of emission: HLX-1 itself (point-like), and diffuse emission peaking in the nuclear region of ESO\,243-49 (Figure~\ref{fig:chandra_contours}). We used {\it{specextract}} to create three spectra and associated response and ancillary response files: for HLX-1 in 2010 and 2022, and for the diffuse emission of ESO\,243-49 (stack of 2010 and 2022 data). For HLX-1, the source extraction region was a circle with a radius of 2\farcs5; for the background, we took 
three circles, each with a radius of 4$^{\prime\prime}$, centred $\approx$7$^{\prime\prime}$ from HLX-1, and located in a suitable way to avoid overlapping with the nuclear region of ESO\,243-49. (The location of three background circles is shown in Figure~\ref{fig:chandra_contours}.) 
For the galaxy emission, we extracted two spectra with a different choice of source region. First, we used an annulus centred on HLX-1, with inner radius of 3$^{\prime\prime}$ and outer radius of 16\farcs5. We centred it at the HLX-1 position because our main goal is to disentangle the origin of the emission seen by {\it{Swift}}/XRT in its monitoring of HLX-1. The choice of a 16\farcs5 radius is again motivated by the comparison with {\it{Swift}}/XRT: the online light-curve tool \citep{evans07,evans09} uses source extraction regions of that radius, for count rate levels comparable to the current (post-2017) level. Second, we extracted a spectrum of the diffuse emission from a circular region of radius 5$^{\prime\prime}$ centred on the nucleus of ESO\,243-49. The advantage of this extraction region is that it minimises photon contamination from HLX-1 coming from the wings of its point spread function (PSF); however, it does not include all the galaxy (Figure~\ref{fig:chandra_hst}) and is therefore a lower limit to its total X-ray emission. Finally, the background region for the diffuse galaxy spectra was taken (in both cases) from an annulus around HLX-1, with inner radius of 20$^{\prime\prime}$ and outer radius of 36$^{\prime\prime}$. 

\subsection{X-ray emission of HLX-1}

The {\it{Chandra}} spectra from the 2010 outburst, together with {\it{XMM-Newton}} and {\it{Swift}} spectra from the same epoch, were analysed and discussed in details in the literature \citep{davis11,servillat11,godet12,soria17}. 
Such investigations found that the spectrum is equally well fitted with a mildly piled-up ($\approx$1150 net counts in 10 ks), standard {\it{diskbb}} model, or slim disk models, or Kerr disk models, or Comptonisation models. In all cases, the best-fitting peak colour temperature of the disk is $\approx$0.22--0.25 keV, and the unabsorbed 0.3--10 keV luminosity is in the $\approx$7--$11 \times 10^{41}$ erg s$^{-1}$ range, depending on the details of the model \citep{davis11,servillat11,godet12}. 

The stacked 2022 spectrum, by contrast, has only 20 raw counts ($\approx$18 net counts): thus, it is impossible to do multi-parameter fits. However, if we assume a spectral model, we can at least estimate its flux and luminosity, to be compared with the luminosity of the optical counterpart. 

In our first scenario, HLX-1 is in the canonical low/hard state with a power-law spectrum with photon index $\Gamma = 1.7$. We fit a {\it{tbabs}}$\times${\it{ztbabs}}$\times${\it{pow}} model, also fixing the Galactic line-of-sight column density at $N_{\rm{H}} = 1.5 \times 10^{20}$ cm$^{-2}$. We obtain a best fit with a Cash statistics of 12.7 over 18 degrees of freedom,  intrinsic $N_{\rm{H}} = 2.2^{+2.9}_{-0.8} \times 10^{22}$ cm$^{-2}$, absorbed 0.3--10 keV flux $F_{\rm{X}} = 5.6^{+3.2}_{-2.2} \times 10^{-15}$ erg cm$^{-2}$ s$^{-1}$, unabsorbed luminosity $L_{\rm{X}} = 1.1^{+0.9}_{-0.5} \times 10^{40}$ erg s$^{-1}$. This solution is physically implausible because it requires much greater intrinsic absorption than inferred in previous observations of this system, at much higher signal-to-noise \citep{servillat11,soria17}. There is also no indication from the evolution of the optical counterpart of a sudden increase in intrinsic $N_{\rm{H}}$ in 2022 compared with the reddening fitted to previous epochs.

In our second scenario, we fit a {\it{tbabs}}$\times${\it{ztbabs}}$\times${\it{pow}} model with fixed intrinsic absorption column density  $N_{\rm{H}} = 5 \times 10^{20}$ cm$^{-2}$ and free power-law photon index and normalisation. This choice of intrinsic $N_{\rm{H}}$ is approximately based on the upper limit of the values fitted to earlier {\it{Chandra}} and {\it{XMM-Newton}} observations \citep{servillat11,soria17}. In fact, considering the degradation of the ACIS-S sensitivity below $\sim$1 keV, choosing intrinsic $N_{\rm{H}} = 0$ would have given the same result. The best fit has a Cash statistics of 14.0/18, a photon index $\Gamma = 0.8^{+0.9}_{-1.0}$, absorbed $F_{\rm{X}} = 7.4^{+7.3}_{-3.4} \times 10^{-15}$ erg cm$^{-2}$ s$^{-1}$, unabsorbed $L_{\rm{X}} = 8.6^{+8.4}_{-3.9} \times 10^{39}$ erg s$^{-1}$.

For our third scenario, we tried thermal emission models. We fit a single temperature (redshifted) blackbody model {\it{tbabs}}$\times${\it{ztbabs}}$\times${\it{zbb}}, also with fixed intrinsic $N_{\rm{H}} = 5 \times 10^{20}$ cm$^{-2}$ and free temperature and normalisation. The Cash statistics is 12.4/18, with a best-fitting $T_{\rm{bb}} = 1.1^{+0.9}_{-0.4}$ keV, absorbed $F_{\rm{X}} = 4.2^{+4.4}_{-1.9} \times 10^{-15}$ erg cm$^{-2}$ s$^{-1}$, unabsorbed $L_{\rm{X}} = 4.9^{+5.1}_{-2.3} \times 10^{39}$ erg s$^{-1}$. Using instead a disk-blackbody model ({\it{tbabs}}$\times${\it{ztbabs}}$\times${\it{diskbb}}), we obtain a Cash statistics of 13.8/18. The upper limit of the peak disk-blackbody temperature is unconstrained ($T_{\rm{dbb}} > 1.3$ keV). The best fit has an absorbed $F_{\rm{X}} = 5.9^{+5.4}_{-3.1} \times 10^{-15}$ erg cm$^{-2}$ s$^{-1}$ and an unabsorbed $L_{\rm{X}} = 6.9^{+6.1}_{-3.5} \times 10^{39}$ erg s$^{-1}$.
A thermal spectrum (blackbody or disk-blackbody) at a luminosity $\lesssim$0.004$\times L_{\rm{Edd}}$, where $L_{\rm{Edd}} \approx 2.5 \times 10^{42}$ erg s$^{-1}$ for a 20,000-$M_\odot$ BH, would obviously be inconsistent with canonical accretion states applied to an IMBH. We still do not know whether HLX-1 does indeed evolve along such states; in fact, we do know that its outbursts are probably not triggered by the thermal-viscous disk instability on which the canonical state transitions are usually defined
\citep{lasota11}.

While the low count statistics in the 2022 dataset preclude a statistically significant distinction between the thermal and power-law models, the power-law description (Scenario 2) is the most consistent with standard accretion physics for a BH at low Eddington ratios ($\lesssim 10^{-2} L_{\rm Edd}$).  Furthermore, Scenario 2 yields the highest unabsorbed luminosity ($L_{\rm X} \approx 8.6 \times 10^{39}$ erg s$^{-1}$) among the statistically acceptable fits.
In the discussion that follows, we will adopt this ``maximum luminosity'' value as our baseline. By utilizing the highest plausible X-ray flux, we ensure that our testing of the irradiated disk hypothesis in Section 5 is conservative; if the observed optical/UV flux is too bright to be explained even by this maximised X-ray input, the irradiated disk model is ruled out regardless of the spectral ambiguities.

Finally, we used Web{\sc{pimms}} to estimate the {\it{Swift}}/XRT count rate (defined as Grade 0--12 on-axis rate for an infinite extraction region) generated by HLX-1 in the three scenarios. Despite the uncertainties in the spectral properties of HLX-1, the predicted rates are similar: $\approx$7.6 $\times 10^{-5}$ ct s$^{-1}$, $\approx$9.4 $\times 10^{-5}$ ct s$^{-1}$, $\approx$7.9 $\times 10^{-5}$ ct s$^{-1}$ for the three scenarios, respectively. Such values are much lower than the typical count rate of several times $10^{-4}$ ct s$^{-1}$ measured by {\it{Swift}}/XRT after 2017 (Figure~\ref{fig:xrt_lc}). This suggests that the observed XRT rate is dominated by the diffuse emission from ESO\,243-49 (Section 3.3).





\subsection{X-ray emission of ESO~243-49}

We expect two types of contribution from the relatively massive spheroidal galaxy ESO\,243-49: i) optically thin thermal plasma emission from diffuse hot gas, typically at the virial temperature; ii) unresolved emission from a population of low-mass X-ray binaries, mostly in the low/hard state, although the brightest ones may reach $\sim$10$^{39}$ erg s$^{-1}$. Based on this prior knowledge, we will model the {\it{Chandra}}/ACIS spectra, and compare the measured flux both with the flux detected by {\it{Swift}}/XRT and with what is theoretically expected from a galaxy the size of ESO\,243-49.

First, we modelled the emission from the larger region (16\farcs5 radius) with a power-law plus thermal plasma model. We fit the 2010 spectrum and 2022 (combined) spectrum, simultaneously. The problem is that in 2010, there is a significant contamination of photons from HLX-1 falling in the wings of the PSF, well outside our exclusion radius of 3$^{\prime\prime}$. In 2020, HLX-1 had $\approx$1150 ACIS-S counts, so even a typical $\approx$2--3 per cent of counts falling outside 3$^{\prime\prime}$ means that about 40 to 60 per cent of the $\approx$50 net counts detected in 2010 in the galaxy extraction region are in fact a contamination from HLX-1. No significant contamination is present in the 2022 dataset because HLX-1 only has $\approx$20 photons; however, the decreased ACIS sensitivity in 2022 makes it impossible to constrain the soft thermal component. In summary, we fitted a {\it{tbabs}}$\times$({\it{apec}}$+${\it{zpow}}), including only line-of-sight extinction and assuming a 50\% contamination in the 2010 flux. The best-fitting model (Cstat $= 187.3/202$\, d.o.f.) has an observed flux $F_{\rm{X,gal}} = 2.2^{+0.9}_{-0.5} \times 10^{-14}$ erg cm$^{-2}$ s$^{-1}$, corresponding to a best-fitting unabsorbed luminosity $\approx$2.7 $\times 10^{40}$ erg s$^{-1}$. The best-fitting split between the two components is $F_{\rm{X,gal,apec}} \approx 0.9 \times 10^{-14}$ erg cm$^{-2}$ s$^{-1}$ (with $kT_{\rm{apec}} = 0.21^{+0.14}_{-0.10}$ keV) and $F_{\rm{X,gal,pow}} \approx 1.3 \times 10^{-14}$ erg cm$^{-2}$ s$^{-1}$ (with $\Gamma = 2.2^{+1.0}_{-1.2}$).  This corresponds to unabsorbed luminosities of $\approx$1.1 $\times 10^{40}$ erg s$^{-1}$ in the thermal plasma component and $\approx$1.6 $\times 10^{40}$ erg s$^{-1}$ in the power-law component. 
Finally, we converted the observed fluxes into {\it{Swift}}/XRT count rates with Web{\sc{pimms}} (separately, for the thermal and power-law components). We predict a best-fitting XRT count rate $\approx$7.3 $\times 10^{-4}$ ct s$^{-1}$, almost independent on the relative split between apec and power-law components. Considering also the uncertainty on the contamination fraction from HLX-1, we can conclude that the emission from ESO\,243-49 contributes $\approx$(5.5--9)$\times 10^{-4}$ ct s$^{-1}$, an order of magnitude more than the contribution from HLX-1 in 2022.




\begin{figure}[t]
    \centering
\hspace{-0.5cm}
    \includegraphics[width=0.48\textwidth]{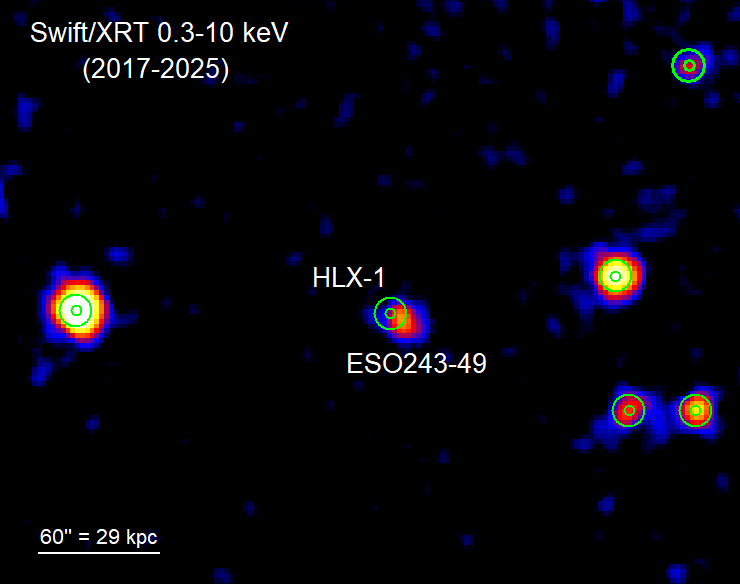}
    \caption{Comparison between the centroid of HLX-1 (and a few neighbouring sources) in the stacked {\it{Swift}}/XRT image built with data from 2008 to 2017 August, including all the outbursts, and in a image built only from the data after the last outburst (2017 September onwards). The false colour map is from the post-2017 data; the green circles map the centroid of the main sources in the 2008--2017 dataset. For each source, the smaller circle has a radius of 2\farcs5, and the larger one a radius of 8$^{\prime\prime}$ (approximate distance between HLX-1 and nucleus of ESO\,243-49). The fitted position of all the sources in the post-2017 map (also those outside the small frame shown here) is consistent with the corresponding position in the pre-2017 map, within 2\farcs5, except for HLX-1, which is significantly offset. This is further evidence that the post-2017 emission is dominated by emission from ESO\,243-49 rather than from HLX-1. 
    }
    \label{fig:swift_offset}
\end{figure}

Second, we repeated the analysis with the smaller source extraction radius (5$^{\prime\prime}$). When power-law index and apec temperature are left free, the best-fitting model (Cstat $= 43.4/51$\, d.o.f.) has an absorbed flux $F_{\rm{X,gal}} \approx 1.5 \times 10^{-14}$ erg cm$^{-2}$ s$^{-1}$, but the relative contribution and fit parameters of the apec and power-law component are largely unconstrained, because of the small number of counts. Fixing instead the apec temperature and the power-law photon index to the values found from the larger extraction region ($kT_{\rm{apec}} = 0.21$ keV, $\Gamma = 2.2$), we obtain a best-fitting model (Cstat $= 44.5/53$\, d.o.f.) with an absorbed flux $F_{\rm{X,gal}} = 1.4^{+0.6}_{-0.4} \times 10^{-14}$ erg cm$^{-2}$ s$^{-1}$ (of which, $F_{\rm{X,gal,apec}} \approx 0.7 \times 10^{-14}$ erg cm$^{-2}$ s$^{-1}$ and $F_{\rm{X,gal,pow}} \approx 0.8 \times 10^{-14}$ erg cm$^{-2}$ s$^{-1}$). The unabsorbed luminosities are $L_{\rm{X,gal,apec}} \approx L_{\rm{X,gal,pow}} \approx 9 \times 10^{39}$ erg s$^{-1}$.
The corresponding count rate in {\it{Swift}}/XRT is $\approx$5 $\times 10^{-4}$ ct s$^{-1}$ (sum of apec plus power-law contributions). We recall that this is the lower limit to what {\it{Swift}}/XRT would measure.

We then looked at the {\it{Swift}}/XRT measurements, to verify the consistency with our {\it{Chandra}} analysis. First, we built two XRT images: one with data until the end of the last outburst (up to MJD 58000), and the other from MJD 58000 onwards (Figure \ref{fig:swift_offset}). The coordinates of all the sources detected in the early-time and late-time maps are identical within $\approx$2\farcs5, except for our target source, in which the post-2017 centroid is offset by $7^{\prime\prime} \pm 1^{\prime\prime}$ to the south west of the location in the pre-2017 map. This is further visual confirmation that after the last outburst, the emission from ESO\,243-49 has been dominant over that from HLX-1. The average net count rate (including the corrections to infinite aperture and for vignetting) of HLX-1 plus ESO\,243-49 since 2017 September is $(5.8 \pm 1.3) \times 10^{-4}$ ct s$^{-1}$ (Figure \ref{fig:xrt_lc}), consistent with the luminosity of the galaxy measured from \textit{Chandra}.

\begin{deluxetable*}{l c c c c c}
\label{tab:optical_ctp}
\digitalasset
\tablewidth{0pt}
\tablecaption{Best-fitting parameters of the UV/optical/IR SED. The model is {\it{redden}}$\times$({\it{zredden$_{\mathrm{h}}$}}$\times${\it{zbb$_{\mathrm{h}}$}}$+${\it{zredden$_{\mathrm{c}}$}}$\times${\it{zbb$_{\mathrm{c}}$}}); best-fitting $\chi^2_\nu = 30.3/26$. The first reddening component was fixed at the line-of-sight value $E(B-V) = 0.011$ mag. Errors are 90\% confidence limits for one interesting parameter. Values in square bracket are linked across all five epochs.} 
\tablehead{
Parameter & \multicolumn{5}{c}{Values}\\
  &  2010  & 2012  & 2013  & 2018  & 2022 \\[-12pt]
}
\startdata\\[-5pt]
$E(B-V)_{\mathrm{h}}$ (mag) & $\left[0.136^{+0.008}_{-0.008}\right]$ & $\left[0.136^{+0.008}_{-0.008}\right]$ & $\left[0.136^{+0.008}_{-0.008}\right]$ & $\left[0.136^{+0.008}_{-0.008}\right]$ & $\left[0.136^{+0.008}_{-0.008}\right]$ \\[5pt]
$T_{\mathrm{bb,h}}$ (K) & $26100^{+900}_{-1000}$ & $28000^{+1000}_{-1000}$ & $26900^{+1300}_{-1300}$ &  $30200^{+1100}_{-1300}$ &  $32000^{+1300}_{-1200}$\\[5pt]
$N_{\mathrm{bb,h}}^a$ $\left(10^{-7}\right)$ & $6.63^{+0.42}_{-0.39}$ & $4.48^{+0.28}_{-0.25}$ & $1.74^{+0.13}_{-0.12}$&  $1.70^{+0.12}_{-0.10}$ & $1.60^{+0.11}_{-0.10}$ \\[5pt]
$E(B-V)_{\mathrm{c}}$ (mag) & $\left[0.005^{+0.030}_{-0.005}\right]$ & $\left[0.005^{+0.030}_{-0.005}\right]$ & $\left[0.005^{+0.030}_{-0.005}\right]$ & $\left[0.005^{+0.030}_{-0.005}\right]$ & $\left[0.005^{+0.030}_{-0.005}\right]$ \\[5pt]
$T_{\mathrm{bb,c}}$ (K) & $\left[4720^{+210}_{-220}\right]$ & $\left[4720^{+210}_{-220}\right]$ & $\left[4720^{+210}_{-220}\right]$ & $\left[4720^{+210}_{-220}\right]$ & $\left[4720^{+210}_{-220}\right]$ \\[5pt]
$N_{\mathrm{bb,c}}^a$ $\left(10^{-7}\right)$ & $\left[0.265^{+0.014}_{-0.012}\right]$  & $\left[0.265^{+0.014}_{-0.012}\right]$ & $\left[0.265^{+0.014}_{-0.012}\right]$ & $\left[0.265^{+0.014}_{-0.012}\right]$  &  $\left[0.265^{+0.014}_{-0.012}\right]$ \\[5pt]
\hline\\[-8pt]
$R_{\mathrm{bb,h}}$ $\left(10^{13}\, {\mathrm{cm}} \right)$ & $1.39^{+0.05}_{-0.04}$ & $0.99^{+0.03}_{-0.03}$ & $0.67^{+0.03}_{-0.02}$ & $0.52^{+0.02}_{-0.02}$ & $0.45^{+0.02}_{-0.02}$ \\[5pt] 
$R_{\mathrm{bb,c}}$ $\left(10^{13}\, {\mathrm{cm}} \right)$ & $\left[8.49^{+0.21}_{-0.20}\right]$ & $\left[8.49^{+0.21}_{-0.20}\right]$ & $\left[8.49^{+0.21}_{-0.20}\right]$ & $\left[8.49^{+0.21}_{-0.20}\right]$ & $\left[8.49^{+0.21}_{-0.20}\right]$\\[5pt]
$L_{\mathrm{bb,h}}$ $\left(10^{40}\, {\mathrm{erg~s}}^{-1} \right)$ & $6.38^{+0.40}_{-0.37}$ &  $4.31^{+0.27}_{-0.24}$  & $1.67^{+0.13}_{-0.12}$ & $1.63^{+0.12}_{-0.09}$ & $1.54^{+0.10}_{-0.10}$  \\[5pt]
$L_{\mathrm{bb,c}}$ $\left(10^{39}\, {\mathrm{erg~s}}^{-1} \right)$ & $\left[2.55^{+0.13}_{-0.12}\right]$  & $\left[2.55^{+0.13}_{-0.12}\right]$ & $\left[2.55^{+0.13}_{-0.12}\right]$ & $\left[2.55^{+0.13}_{-0.12}\right]$  &  $\left[2.55^{+0.13}_{-0.12}\right]$ \\[5pt]
\enddata
\tablecomments{$^a$: defined as $L_{39}/D^2_{10}$, where $L_{39}$ is the bolometric luminosity in units of $10^{39}$ erg s$^{-1}$, and $D_{10}$ is the luminosity distance to the source in units of 10 kpc (here, $D_{10} = 9800$).}
\end{deluxetable*}

Finally, we compare the X-ray luminosity of ESO\,243-49 as observed by {\it{Chandra}} and {\it{Swift}} with the theoretical expectations for a galaxy of that mass. From the results of \cite{webb17}, we obtain an apparent brightness of ESO\,243-49 (bulge plus disk) $m_{\rm{H}} \approx 11.22$ (Vega) mag in the WFC3-IR F160W band (approximately coincident with the Johnson H band). Correcting for Galactic extinction and using a luminosity distance of 98 Mpc gives an absolute magnitude $M_{\rm{H}} \approx -23.75$ mag. Then, the stellar mass 
$M_{\ast} = (M/L_{\rm{H}}) \times 10^{(23.75+3.36)/2.5} 
= (M/L_{\rm{H}}) \times 7.0 \times 10^{10} M_\odot$, where the mass-to-light ratio $M/L_{\rm{H}}$ is in solar units, and $L_{\rm{\odot,H}} \approx 3.36$ mag.
For the stellar age and metallicity of bulge and disk in ESO\,243-49 \citep{comeron16,webb17}, a Salpeter IMF \citep{salpeter55} suggests an average $M/L_{\rm{H}} \approx 1.15 M_\odot/L_{\rm{\odot,H}}$. A Kroupa IMF \citep{kroupa02} yields half that value. A diet-Salpeter IMF \citep{bell01,bell03} gives $M/L_{\rm{H}} \approx 0.69 M_\odot/L_{\rm{\odot,H}}$ for the observed $B-V \approx 0.71$ mag (from NED, dereddened). 
Taking for example the diet-Salpeter value, $M_{\ast} \approx 4.8 \times 10^{10} M_\odot$. Stellar-mass values in the range of 2.5--7 $\times 10^{10} M_\odot$ depending on the choice of IMF are also found in the modelling of \cite{mapelli13b}; a mass of $M_{\ast} \approx 6 \times 10^{10} M_\odot$ was suggested by \cite{soria10}, based on optical colours and luminosities.

Knowing that the early-type galaxy ESO\,243-49 has a stellar mass of a few $\times 10^{10} M_\odot$, we can estimate its expected X-ray luminosity from discrete sources and diffuse emission.
Low-mass X-ray binaries contribute an average $L_{\rm{XRB}} \approx 10^{(29.25 \pm 0.10)} (M_{\ast}/M_\odot)$ in the 0.5--8 keV band \citep{lehmer19}. Thus, for a diet-Salpeter IMF, $L_{\rm{XRB}} \approx (0.9\pm0.2) \times 10^{40}$ erg s$^{-1}$ in that band, or $L_{\rm{XRB}} \approx (1.1\pm0.2) \times 10^{40}$ erg s$^{-1}$ in the 0.3--10 keV band. The X-ray luminosity of the hot gas within the virial radius (``hot circumgalactic medium'') in early-type galaxies is also a function of $M_{\ast}$. A recent galaxy survey based on {\it{Spektrum Roentgen Gamma}} eROSITA data \citep{zhang24} suggests that for $M_{\ast} \approx 5.0 \times 10^{10} M_\odot$, $L_{\rm{CGM}} \approx (0.7\pm0.3) \times 10^{40}$ erg s$^{-1}$ in the 0.5--2 keV band, or $L_{\rm{CGM}} \approx (1.0\pm0.4) \times 10^{40}$ erg s$^{-1}$ in the 0.3--10 keV band.

In conclusion, the predicted luminosity of X-ray binaries plus hot gas in a galaxy with the stellar properties of ESO\,243-49 is in remarkably good agreement with the measured X-ray luminosity in the power-law and thermal plasma components from the {\it{Chandra}} and {\it{Swift}} data. This implies that the post-2017 plateau in the observed {\it{Swift}}/XRT flux is the approximately constant level of galaxy emission.

\section{Optical/UV results}

\subsection{Evolution of the optical counterpart of HLX-1}

Earlier work \citep{soria17} noted the dimming and reddening of the optical counterpart from 2010 to 2013. Here, we monitor and interpret the long-term evolution with additional {\it{HST}} data from 2018 and 2022. Our new measurements confirm (Table \ref{tab:hst_mag_results}) a declining trend, much stronger in the bluer bands. For instance, in the far-UV F140LP band, HLX-1 became $\approx$1.7 mag fainter from 2010 to 2022, while in the near-IR F775W band it only declined by $\approx$0.6 mag over the same time span. 


We converted the observed {\it{HST}} flux-density datapoints to {\sc{xspec}} format, with the {\sc{ftools}} task {\it{ftflx2xsp}}, and fitted the spectral energy distributions with simple phenomenological models, using $\chi^2$ minimisation. When doing this, we added a systematic error of 5\% in {\sc{xspec}}, justified by the additional uncertainty introduced by the conversion process, which assumes rectangular filters and a flat spectrum in each band. 

We find that each epoch is well fitted with two single-temperature blackbody components: in {\sc{xspec}}, the model is {\it{redden}}$\times$({\it{zredden$_{\mathrm{h}}$}}$\times${\it{zbb$_{\mathrm{h}}$}}$+${\it{zredden$_{\mathrm{c}}$}}$\times${\it{zbb$_{\mathrm{c}}$}}), where the ``h'' and ``c'' suffixes indicate the hotter and colder blackbody, respectively, and the redshift was fixed at $z = 0.0224$. More complex models ({\it{e.g.}}, replacing the hotter blackbody with an irradiated disk model) do not provide any statistical advantage. The first reddening term was fixed at the line-of-sight Galactic value $E(B-V) = 0.011$ mag \citep{schlafly11}; the intrinsic reddening terms (one for the hotter and one for the colder component) were left free but each one locked between all five epochs. Initially, we left normalisation and temperature free to vary for both blackbody components. This showed that both temperature and normalisation of the colder blackbody are consistent with a constant value across the five epochs. Finally, we fitted a model with 40 datapoints (corresponding to the 40 entries of Table \ref{tab:hst_mag_results}) and 14 free parameters (two intrinsic reddenings; one temperature and one normalisation for the colder component; one temperature and one normalisation per epoch for the hotter component).

\begin{figure}[t]
\hspace{-0.5cm}
    \includegraphics[width=0.35\textwidth, angle=270]{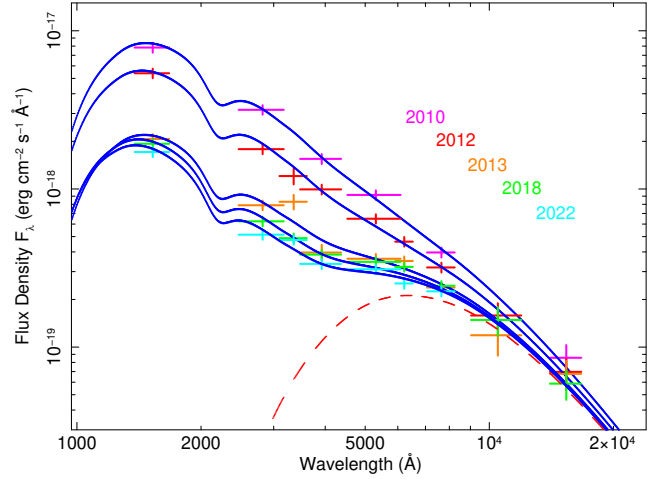}
\vspace{0.7cm}
    \caption{Unfolded SED of the UV/optical/IR band, from the {\it{HST}} observations, fitted with a double thermal model (see Table \ref{tab:hst_mag_results} for the datapoint values and Table \ref{tab:optical_ctp} for the best-fitting parameters). Each blue line represents the (unfolded) best fitting model for one observing epoch: from top to bottom, 2010, 2012, 2013 2018 and 2022. Datapoints along each of the model fits are displayed in different colours (magenta for 2010, red for 2012, orange for 2013, green for 2018 and cyan for 2022). The dashed red line represents the redder of the two blackbody components, assumed constant in our modelling. }  
    \label{fig:hst_allyears_lambda}
\end{figure}

\begin{figure}[t]
\hspace{-0.5cm}
    \includegraphics[width=0.35\textwidth, angle=270]{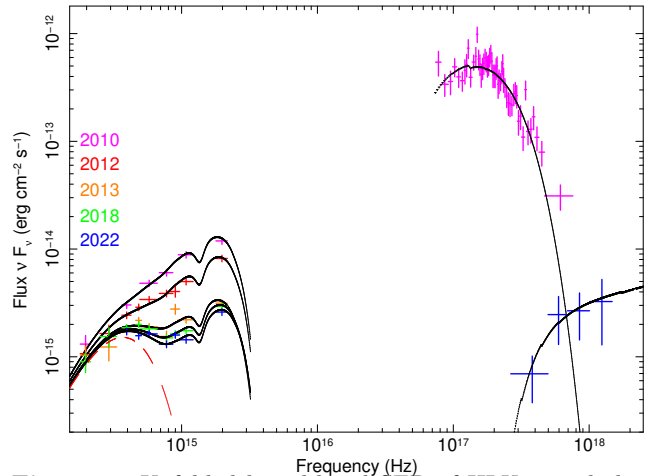}
\vspace{0.7cm}
    \caption{Unfolded broad-band SED of HLX-1, including the five {\it{HST}} observations (magenta datapoints for 2010, red for 2012, orange for 2013, green for 2018 and blue for 2022) and the two {\it{Chandra}} observations (magenta datapoints for 2010, blue for 2022). The {\it{HST}} data have been modelled with a double blackbody component (Table \ref{tab:optical_ctp}); the dashed red line represents the redder of the two components, assumed constant. The {\it{Chandra}} data are fitted by a disk blackbody in 2010 and a power-law in 2022 (Section 3.2).
    }
    \label{fig:sed_chandra}
\end{figure}

We found (Figure \ref{fig:hst_allyears_lambda} and Table \ref{tab:optical_ctp}) that the decreasing trend of the bluer component continues, although slower than in the early years. This is further evidence that the blue/UV emission comes from gas somehow associated with the X-ray activity, rather than from a young star cluster. The standard interpretation has been that the bluer component comes from the irradiated accretion disk, while the redder component traces an old stellar population, probably the host star cluster or remnant of a satellite galaxy \citep{mapelli13b,soria17}. 

The irradiated disk interpretation of the blue/UV emission is perfectly consistent with the 2010 data, an epoch when a plausible X-ray reprocessing fraction $f_{\rm{irr}} \sim$ a few $\times 10^{-3}$ suffices to fit the optical data \citep{mapelli13b,farrell14,soria17}. When HLX-1 was in an intermediate X-ray state, in 2012--2013, the irradiated disk model was still applicable but required an unusually high reprocessing fraction $f_{\rm{irr}} \sim$ a few $\times 10^{-2}$ \citep{soria17}. Now, we show (Figure \ref{fig:sed_chandra}) that after the last outburst, over the 2018--2022 period, the X-ray luminosity is lower than (or at most similar to) the optical/UV luminosity: $L_{\rm{X}} \lesssim 10^{40}$ erg s$^{-1}$ (Section 3.2 and Figure \ref{fig:xrt_lc}), compared with $L_{\rm{bb,h}} \approx 1.5$--$1.6 \times 10^{40}$ erg s$^{-1}$ (Table \ref{tab:optical_ctp}). This implies that the irradiated disk model is inconsistent with the combined {\it{HST}} and {\it{Chandra}} data, at least for post-2017 observations. 

\begin{figure}[t]
    \hspace{-0.5cm}
    \includegraphics[width=0.35\textwidth, angle=270]{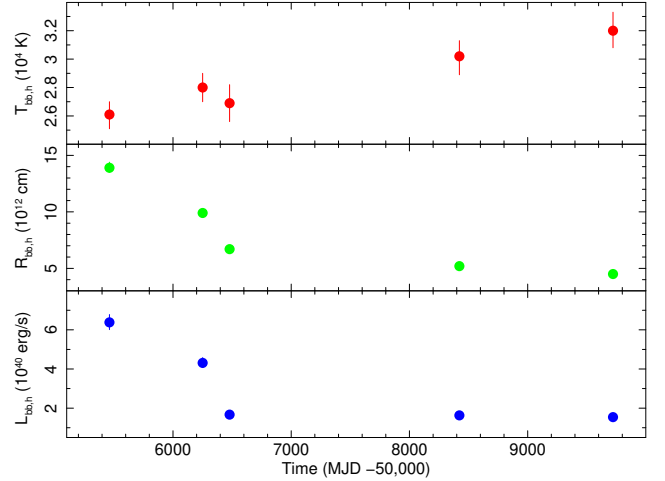}
\vspace{0.5cm}
    \caption{Evolution of the best-fitting blackbody temperature, radius and luminosity of the UV/optical blackbody component, as a function of time of observation. (See also Table \ref{tab:optical_ctp}.)
    }
    \label{fig:hst_components1}
\end{figure}

\begin{figure}[t]
    \hspace{-0.5cm}
    \includegraphics[width=0.35\textwidth, angle=270]{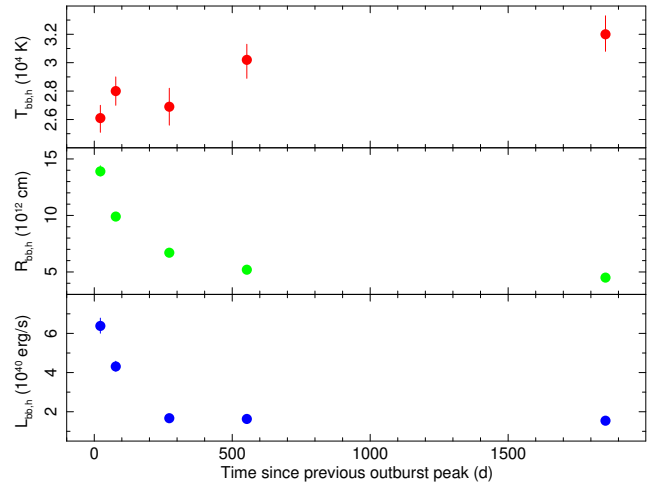}
\vspace{0.5cm}
    \caption{As in Figure \ref{fig:hst_components1}, but with temperature, radius and luminosity plotted as a function of time since the previous X-ray outburst peak.
    }
    \label{fig:hst_components2}
\end{figure}

Another clear trend of the optical emission is the decrease in the characteristic area (Figures \ref{fig:hst_components1},\ref{fig:hst_components2}). The best-fitting blackbody radius of the bluer component was $R_{\rm{bb,h}} \approx 1.4 \times 10^{13}$ cm in 2010 (consistent with the outer disk size $R_{\rm{bb,h}} \sim 2 \times 10^{13}$ cm inferred by \cite{soria17} from an irradiated disk model). By 2022, it had decreased by a factor of 3 ($R_{\rm{bb,h}} \approx 4.5 \times 10^{12}$ cm). In a standard, irradiated disk model, if the outer ring of the optically thick part of the disk decreases as observed, we would expect a decrease of the optical/UV luminosity (mostly due to X-ray reprocessing) steeper than the decrease in X-ray luminosity ({\it{i.e.}}, a smaller radiative power from the central engine compounded by a smaller fraction of that X-ray luminosity intercepted by the outer disk), not in agreement with what is observed. 

Third, we find a mild but significant increase in blackbody temperature (Table \ref{tab:optical_ctp} and Figures \ref{fig:hst_components1},\ref{fig:hst_components2}) over the years (from $\approx26,000$ K in 2010 to $\approx32,000$ K in 2022), associated with the decline in luminosity and radius. This is another finding difficult to reconcile with an irradiated disk model, in which the outer disk temperature typically decreases when the X-ray luminosity drops.

\subsection{A far-UV emitting structure behind HLX-1}
An intriguing additional feature of the HLX-1/ESO\,243-49 system is the 
presence of an extended far-UV source (projected) between HLX-1 and the galactic nucleus \citep{wiersema10,mapelli13b,soria13,webb17}. The heliocentric recession speed of the brightest part of this region is $\approx$9170 km s$^{-1}$ ($z \approx 0.0306$), that is about 2000 km s$^{-1}$ more than the recession speed tentatively attributed to HLX-1 on the basis of a single emission line \citep{wiersema10,soria13}. For this discrepancy, the far-UV emitter is generally dismissed as a background structure obviously unrelated to HLX-1. 
We will discuss later (Section 5.3) whether there are scenarios in which HLX-1 and the far-UV-emitting structure could be physically related. First, we quantify the far-UV luminosity of the structure, using {\it{HST}}/ACS and {\it{Swift}}/UVOT.  

\begin{figure}[t]
    \centering
    \begin{minipage}{0.47\textwidth}
        \centering
        \includegraphics[width=\linewidth]{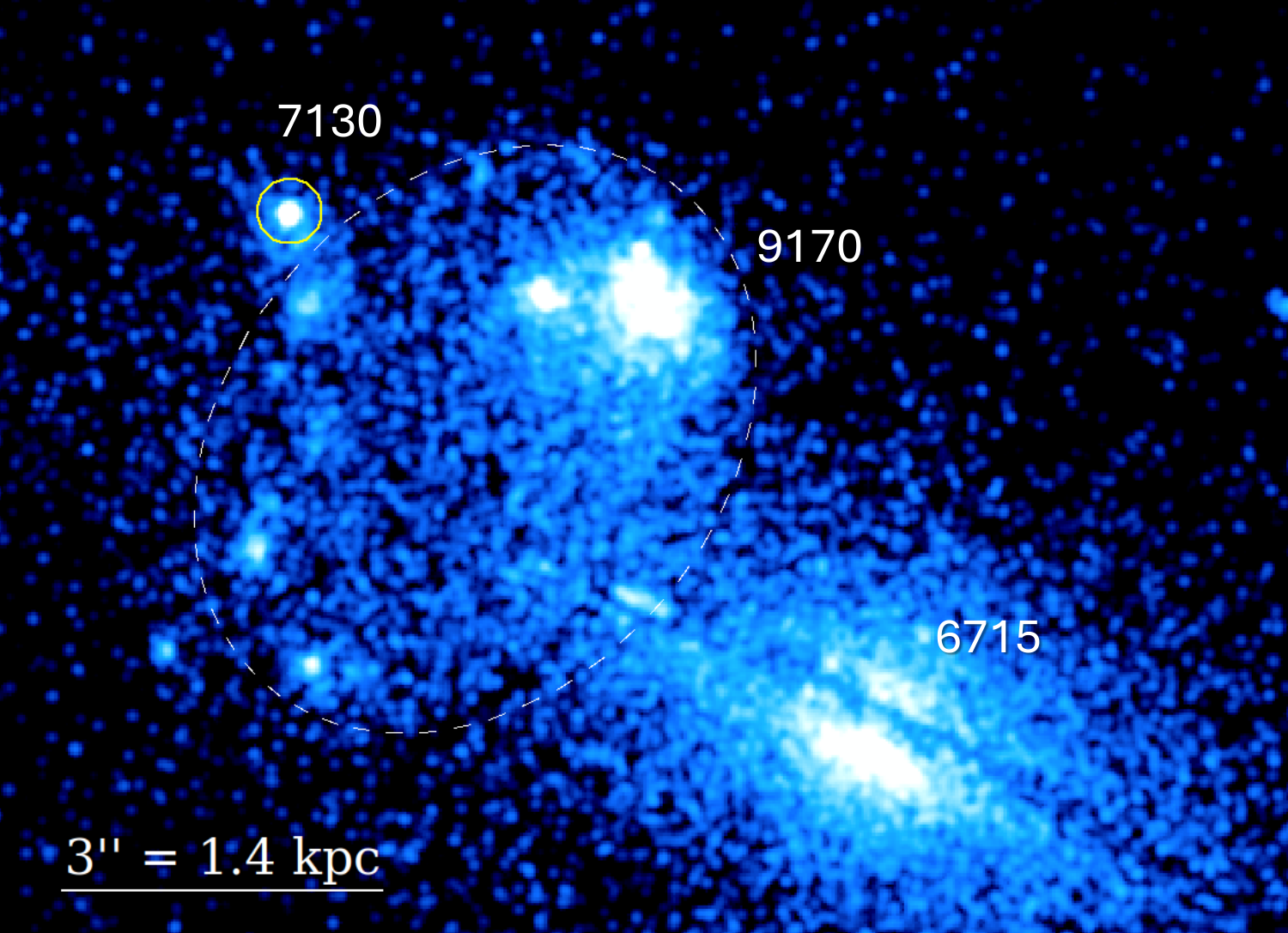}
    \end{minipage}
    \hfill
    \begin{minipage}{0.47\textwidth}
        \centering
        \includegraphics[width=\linewidth]{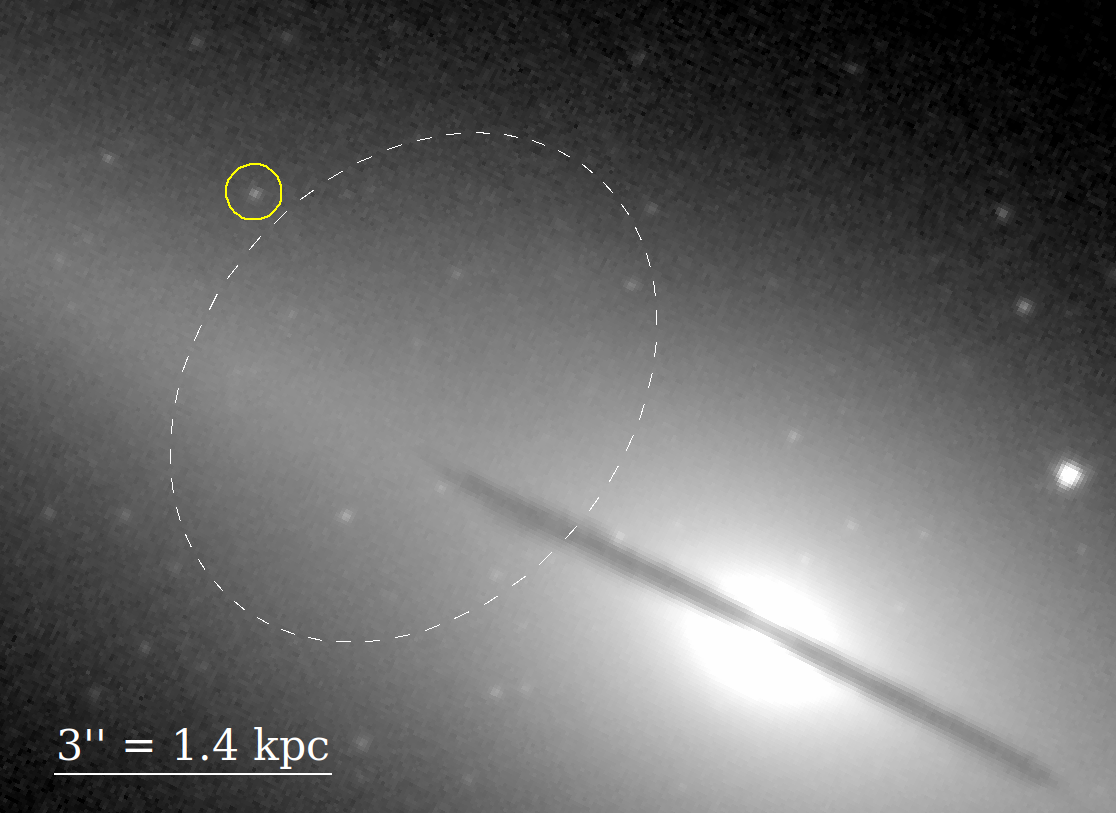}
    \end{minipage}
    \caption{Top panel: {\it{HST}}/ACS image in the far-UV F140LP band (centered at $\lambda \approx 1500$\,\AA), built from a stack of all five observations (Table \ref{tab:hst_mag_results}); north is up and east to the left. HLX-1 is circled in yellow. The stacked image highlights the extended elliptical region of far-UV emission (size of $\approx$6$^{\prime\prime} \times 4^{\prime\prime}$) between HLX-1 and the nucleus of ESO\,243-49, noted in earlier studies \citep{wiersema10,mapelli13b,soria13,webb17}. The labels indicate the heliocentric recession speeds $cz$ (in km s$^{-1}$) of the three main structures in the field: the nucleus of ESO\,243-49, the brightest part of the far-UV emitter, and HLX-1 (the last one based on a single emission line tentatively identified as H$\alpha$; \citep{wiersema10,soria13}). Bottom panel: {\it{HST}}/WFC3 image of the same field in the F555W band; the dashed white ellipse guides the eye to the approximate extent of the far-UV emitting region. No stellar structure distinct from ESO\,243-49 is visible in this or any other optical image.
    }
    \label{fig:nebula_hst}
\end{figure}

We aligned and combined all five {\it{HST}}/ACS observations in the F140LP band (Figure \ref{fig:nebula_hst}, top panel). From the deep combined image, we show that the far-UV emission comes from an elliptical region, with major axes of $\approx$6$\arcsec \times 4\arcsec$ ($\approx$3.7 $\times$ 2.5 kpc at $z \approx 0.0306$), and the brightest emission mostly located around the outer rim. HLX-1 is at the edge of this region. A faint detection of the brightest UV cluster is possible also in the {\it{HST}}/WFC3 UVIS F300X band, but the source is all but invisible in redder continuum bands (Figure \ref{fig:nebula_hst}, bottom panel). The UV emission is detectable due to the much lower background levels, while optical emission is likely present but apparently falls below the detection threshold against ESO 243-49's old stellar population light.

We measured the net count rate from the 6$\arcsec \times 4\arcsec$ region (not including the emission from HLX-1 itself) for each of the five observations, then computed the average count rate, and converted it to a flux density using the STScI's ACS zeropoints calculator. We obtain an absorbed flux density $F_{\lambda{\mathrm{,F140LP}}} = (5.1 \pm 0.5) \times 10^{-17} \; \text{erg~cm}^{-2} \text{~s}^{-1} \text{\AA}^{-1}$.  
Correcting only for line-of-sight absorption ($A_{\mathrm{F140LP}} \approx 2.65 A_V \approx 0.095$ mag), and assuming a luminosity distance $d_{\mathrm{L}} = 135 \; \text{Mpc}$ ($z \sim 0.0306$; \citealt{webb17}), we derive a luminosity density $L_{\lambda{\mathrm{,F140LP}}} = (1.2\pm0.1) \times 10^{38} \;  \text{erg~s}^{-1} \text{\AA}^{-1}$.

{\it{Swift}}/UVOT also shows excess UV emission from the region between HLX-1 and the nucleus of ESO\,243-49, as already noted from the first {\it{Swift}} observations \citep{soria10}. Here, we refine and update those earlier flux estimates. We downloaded all UVOT observations of the field from the HEASARC archive, reduced them, and created merged image files for each of the six UVOT filters. The excess is statistically significant only for the shortest-wavelength band ($uvw2$), centered around 1928 \AA (Figure \ref{fig:uvw2_vlt}). 
We quantified the detection significance of the $uvw2$ emission by exploiting ESO\,243-49's approximately symmetrical shape around its mid-plane. We divided the galaxy into four rectangular sectors of approximately equal area. We then measured the count rate in the north-eastern sector, and used the north-western and south-western sectors to define the background level of emission from ESO\,243-49. The resulting null probability is lower than $10^{-5}$, confirming the significance of the far-UV source near HLX-1.

We estimated the \textit{uvw2} flux density via relative photometry, by choosing suitable nearby comparison sources with known \textit{uvw2} measurements. Specifically, we used the following three flux reference sources: PDF J011048.0$-$460133 (a galaxy at $z \approx 0.298$); PDF J011053.2$-$460701 (a galaxy at $z \approx 0.102$); and PDF\,J011009.6$-$460546 (a galaxy at photometric $z \approx 0.18\pm0.05$). For all three sources, UV fluxes in the \textit{uvw2} band are available in the Swift/UVOT Serendipitous Source Catalog \citep{yershov14} and in the XMM-OM Serendipitous Source Survey Catalogue \citep{page12}, on Vizier. 
Thus, we obtain a flux density $F_{\lambda,uvw2} = (1.98 \pm 0.09) \times 10^{-17} \; \text{erg~cm}^{-2} \text{~s}^{-1} \text{\AA}^{-1}$,
which corresponds to an AB magnitude of $m_{uvw2} = (22.92 \pm 0.05)$ mag. 
Correcting only for the line-of-sight absorption $A_{uvw2} \approx 3.09 A_V \approx 0.11$ mag \citep{schlafly11} ({\it{i.e.}}, neglecting the local dust extinction in the star-forming region), and assuming a luminosity distance $d_{\mathrm{L}} = 135 \; \text{Mpc}$ (as opposed to $d_{\mathrm{L}} = 98 \; \text{Mpc}$ assumed for ESO\,243-49 and HLX-1), we infer $L_{\lambda,uvw2} = (4.8\pm0.2) \times 10^{37} \;  \text{erg~s}^{-1} \text{\AA}^{-1}$. 

\section{Discussion}

\subsection{TDE outflow or irradiated disk?}

\label{sec:temp_evol_bbcomps}

We have shown (Section 4.1) that the irradiated disk model faces a critical physical inconsistency when we quantify the reprocessing efficiency required by the late-time optical/UV luminosity. In a standard X-ray binary geometry, the optical emission arises from the outer accretion disk intercepting and re-emitting X-rays from the compact central engine. The efficiency of this process is strictly limited by geometry: the fraction of X-rays intercepted is determined by the solid angle subtended by the disk as seen from the corona. Even accounting for significant disk flaring, canonical models predict a reprocessing fraction $f_{\rm irr} \equiv L_{\rm opt}/L_{\rm X}$ typically in the range of $10^{-3}$ to roughly $5 \times 10^{-2}$ \citep[e.g.,][]{dubus99,gierlinski09}. However, the 2022 dataset reveals a stark contradiction. With an unabsorbed X-ray luminosity of $L_{\rm X} \approx (4.9\text{--}8.6) \times 10^{39}$ erg s$^{-1}$ (Section 3.3) and a ``blue'' thermal component luminosity of $L_{\rm bb,h} \approx 1.5 \times 10^{40}$ erg s$^{-1}$ (Table 3), the required reprocessing fraction is $f_{\rm irr} \approx 1.8\text{--}3.1$. A value of $f_{\rm irr} > 1$ is physically impossible for a passive disk, as it implies the re-processed 
output exceeds the radiative input, violating the conservation of energy. 
To rescue the X-ray irradiation model, one would have to postulate that we are underestimating the intrinsic X-ray luminosity by a factor of $\sim$100 (to bring $f_{\rm irr}$ down to physically plausible levels of $\sim$0.02). This would require either an implausible bolometric correction ({\it{e.g.}}, essentially all energy emitted above 10 keV) or extreme anisotropic beaming where the disk sees a bright X-ray source that is effectively hidden from the observer. Given the lack of spectral evidence for such extreme obscuration or beaming, we conclude that the late-time optical/UV emission cannot be driven by irradiation. Instead, it must originate from a distinct, self-luminous component, consistent  either with the cooling photosphere of a tidal disruption event (TDE) outflow, or with viscous energy dissipation in the TDE-generated outer disk.

The decrease in radius and luminosity accompanied by a slight increase in blackbody temperature are also typical properties of the optical/UV component in TDEs ({\it{e.g.}}, Figure 2 of \citealt{2020SSRv..216..124V}; Figure 8 of \citealt{2021ApJ...908....4V}). Moreover, characteristic temperatures $\approx$3 $\times 10^4$ K are exactly what is observed in TDEs over a wide range of BH masses and evolution times. The decoupling of optical/UV and X-ray luminosities is now being observed in other TDEs at very late times. For example, in ASASSN-14li \citep{brown17}, AT2018fyk \citep{wevers21} and XMMSL2\,J140446.9$-$251135 \citep{saxton25}, the X-ray emission dropped drastically after a few 100 days, while the optical/UV continuum faded much more slowly. The detection of a late-time optical/UV luminosity plateau in HLX-1 (Figures \ref{fig:sed_chandra} and \ref{fig:hst_components1}) provides the clearest analogy with typical TDEs \citep{mummery24}. The plateau occurs at a bolometric blackbody luminosity $L_{\rm bb,h} \approx 1.5 \times 10^{40}$ erg s$^{-1}$, or at a frequency-specific luminosity $\nu L_{\nu} \approx 2 \times 10^{39}$ erg s$^{-1}$ for $\nu \approx 6 \times 10^{14}$ Hz. A simple extrapolation of the empirical scalings between plateau luminosities and BH masses in TDEs \citep{mummery24,alush25} suggests that HLX-1 is consistent with a TDE on a $10^{4} M_\odot$ BH.


Late-time optical/UV plateaus have been explained both within a viscous disk scenario, and within a cooling envelope scenario. For an accretion disk, the plateau represents a balance between disk cooling and disk spreading \citep{mummery24,mummery26}. In fact, we did not observe evidence of optical/UV cooling in HLX-1; for this reason, we favour the outflow scenario. We suggest that the optical/UV continuum comes
from the photosphere (Thomson scattering optical depth $\approx$1) of an expanding outflow driven by the transient accretion episode \citep{metzger16,roth16,dai18,2018ApJ...867...20C,bu22,qiao25}. As the outflow expands, its density decreases, its photospheric radius shrinks into a hotter part of the outflow, causing a slight temperature increase, but still within the $\approx$20,000--40,000 K range; this characteristic temperature is ultimately related to the hydrogen ionisation threshold, which determines the location of the scattering photosphere. Similar physics explains the optical/UV emission in expanding nova envelopes \citep{cunningham15}. In both classes of sources, soft X-ray photons are injected from a central source into the inner part of the envelope, and emerges reprocessed as UV radiation. For TDEs, the main source of power injected into the reprocessing envelope is accretion luminosity, but kinetic power of fast winds from the inner disk, and stream collimation and collision shocks may also be important; for example, wind kinetic power is comparable to radiative power in the super-Eddington accretion regime.

\begin{figure}[t]
\hspace{-0.5cm}
    \includegraphics[width=0.35\textwidth, angle=270]{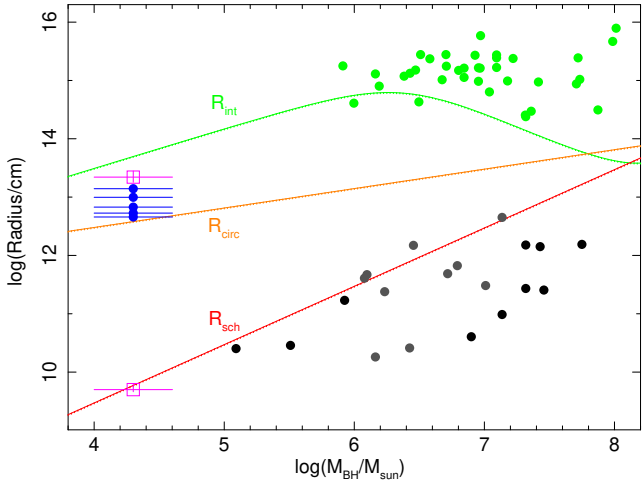}
\vspace{1.0cm}
    \caption
    {Comparison of the characteristic radii inferred for HLX-1 with those measured in thermal TDEs, as a function of BH mass (plot adapted from Figure\ 8 of \citealt{gezari21}). The magenta datapoints are $R_{\rm{out}}$ and $R_{\rm{in}}$ in 2010, inferred from an irradiated disk model by \cite{soria17}, and we have also assumed the mass range $M_{\rm{BH}} = 2^{+2}{-1} \times 10^4 M_\odot$ from the same reference. The blue datapoints are the blackbody radii of the blue/UV component measured from the five {\it{HST}} observations discussed in this work (Table \ref{tab:optical_ctp}). The other datapoints, from \cite{gezari21}, are the characteristic thermal emission radii of optical/UV-selected TDEs (in green), of X-ray-selected TDEs (in black), and of the soft X-ray component of optical/UV-selected TDEs (in grey). The three theoretical curves are the radius of the self-intersecting stream ($R_{int}$, from \cite{Dai_2015}, in green), the circularisation radius ($R_{circ}$, in orange) and the Schwarzschild radius ($R_{\rm{sch}}$, in red).  
    }
    \label{fig:gezari_extension}
\end{figure}

Optical/UV observations of TDEs suggest photospheric radii similar or larger than the self-intersection radius of the accretion stream \citep{gezari21}, much larger than the circularisation radius or the tidal disruption radius. 
By comparison, the blackbody radius of the UV/optical emission in HLX-1 is in between the self-intersection radius and the circularisation radius (Figure \ref{fig:gezari_extension}), if we assume that the thermal emission radius inferred from the X-ray spectrum is a reliable indicator of a BH mass $\sim$10$^4 M_\odot$. The optical/UV luminosity ($\sim$10$^{40}$ erg s$^{-1}$) is also three orders of magnitude lower than what is observed in most TDEs, which is consistent with an IMBH event; see for example the scaling relations between luminosity and BH mass in \cite{mummery24}. We also need to consider that HLX-1 has been observed in the optical/UV only since 2010, already in decline, but the X-ray source had already been detected in 2004.

While the optical/UV emission comes from an outflow photosphere, we also see direct X-ray emission. This indicates that we are looking at the system face-on, along a low-density polar funnel, where some of the X-ray emission from the inner accretion disk can freely escape with minimal absorption \citep{dai18}. 
This outflow geometry naturally explains how direct X-ray emission can coexist with optical/UV signatures from outflows.
The low absorption column density ($N_{\rm{H}} \lesssim 10^{21}$ cm$^{-2}$) usually inferred for the X-ray spectrum \citep{farrell09,servillat11,godet12,soria17} is consistent with this interpretation.

A partial TDE scenario was invoked to explain the cycle of X-ray outbursts seen until 2017 \citep{godet14}; in this case, each periastron passage (with tidal stripping) of the donor star could have contributed to the feeding of the scattering envelope. However, we suggest that the activity of HLX-1 could have been triggered by a single, full TDE shortly before 2004, and that the recurrent outbursts were caused by instabilities in the inner accretion flow: for example a radiation pressure instability \citep{wu16}. In this scenario, the mass fallback rate has been decreasing steadily after the TDE ({\it{e.g.}}, $\propto t^{-5/3}$), but the emerging X-ray luminosity is not a direct tracer of the fallback rate.

\begin{figure}[t]
    \centering
    \begin{minipage}{0.47\textwidth}
        \centering
        \includegraphics[width=\linewidth]{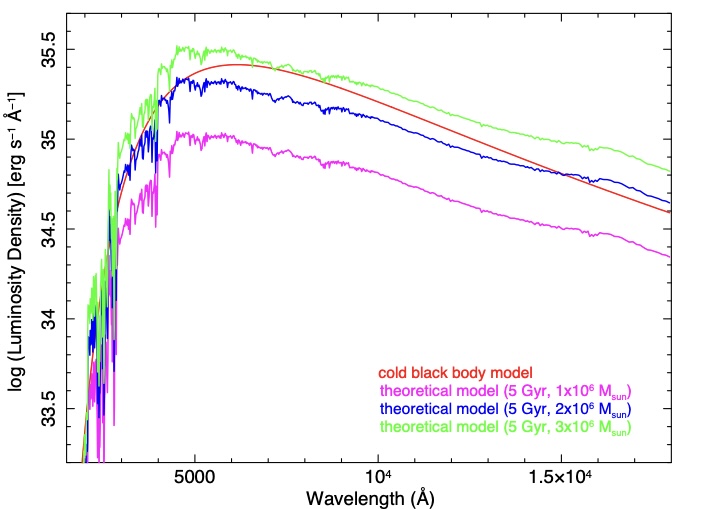}
    \end{minipage}
    \hfill
    \begin{minipage}{0.47\textwidth}
        \centering
        \includegraphics[width=\linewidth]{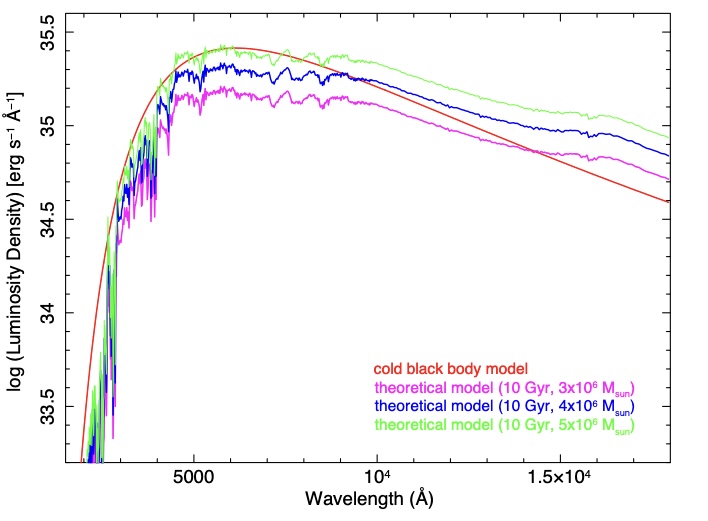}
    \end{minipage}
    \caption{Comparison between the luminosity density of the best-fitting, constant, cool blackbody component in our model (red line; parameters from Table \ref{tab:source_radii}) and those of old globular clusters across a range of masses and ages. In the top panel, we show the theoretical luminosity densities of a 5-Gyr-old cluster with masses of 1, 2 and 3 $\times 10^6 M_\odot$ (magenta, blue and green spectra, respectively). In the bottom panel, we show the same for a 10-Gyr-old cluster with masses of 3, 4 and 5 $\times 10^6 M_\odot$. All cluster spectra were computed with {\sc{starburst99}} \citep{leitherer99,leitherer14}, assuming solar metallicity and instantaneous star formation with Kroupa IMF. }
    \label{fig:cluster_vs_starburst99}
\end{figure}

Recent calculations of the time-dependent evolution of an accretion disk formed by a TDE (with mass injected at the circularisation radius) appear to support the oscillation scenario \citep{guo26}. In such simulations, the X-ray luminosity was found to decrease steadily in the initial super-Eddington phase of mass fallback; then, when $\dot{M} \sim \dot{M}_{\rm{Edd}}$, the flow goes through a limit cycle instability triggered by radiation pressure, with repeated oscillations of the output luminosity, with an increasing recurrence timescale (consistent with what is observed in HLX-1), function of the viscosity parameter in the disk and the injection radius. Finally, when $\dot{M}$ has decreased below a characteristic threshold, the inner disk becomes gas-pressure dominated and stable, and the instability is suppressed \citep{guo26}. 
Other theoretical investigations of accretion after a TDEs have also found an epoch of recurrent X-ray flares, lasting several years, during the smooth decline of the fallback rate \citep{shen14,piro25}. Such studies confirm the prediction of an increasing waiting time between X-ray outbursts, because the gradually decreasing accretion rate causes the disk to cycle more slowly between states. 

Theoretical predictions for the recurrence timescale and luminosity amplitude of the radiation pressure instability are functions of the BH mass, the outer disk radius, the Eddington ratio, the viscosity parameter $\alpha$ and the functional dependence of the stress tensor on gas pressure and radiation pressure \citep{janiuk17,guo26}. For IMBHs with $M \sim 10^4 M_\odot$, there is a plausible range of such parameters that produce X-ray oscillations with an amplitude $>$100, peak luminosity around $L_{\rm{Edd}}$ and recurrent timescales of a few 100 days \citep{wu16,janiuk17,guo26}, consistent with the observed X-ray variability of HLX-1. In this scenario, HLX-1 entered its last phase of evolution (the end of the disk instability phase) after its last outburst in 2017. 
A detailed application of TDE instability models to fit the X-ray light-curve of HLX-1 is left to follow-up work.
As a comparison, the TDE XMMSL2\,J140446.9$-$251135 showed three X-ray flares over a time interval of $\approx$1500 days \citep{saxton25}, and is a plausible candidate for this type of TDE-related disk instability, although the alternative interpretation of a partial TDE is also viable.


Finally, if the optical line emission comes from the outer, optically thin part of an outflow \citep{metzger16,roth18,zhang24b}, rather than from the disk surface, the line centroid may not be an accurate tracer of the systemic velocity of the BH. Specifically, optical lines tend to have a blueshifted peak (due to Doppler effects) and a broader redshifted wing due to photon scattering in the expanding envelope \citep{roth18,nicholl20}. 
In low signal-to-noise spectra, such as those of HLX-1 from 2009 and 2012 used for the emission line detection \citep{wiersema10,soria13}, only the line peak could be used for the profile fit. Therefore, the fitted recession speed of $\approx$7130 km s$^{-1}$ could be only a lower limit to the true recession speed. At $\approx$7130 km s$^{-1}$, HLX-1 already has an $\approx$420 km s$^{-1}$ radial-velocity discrepancy from the systemic velocity of ESO\,243-49, which makes it plausible that it is not bound to the galaxy \citep{soria13}. If the line peak is affected by an additional Doppler shift $\gtrsim$10$^3$ km s$^{-1}$, we cannot rule out that HLX-1 might be associated to the UV-bright star-forming dwarf at $v \approx$9200 km s$^{-1}$ rather than to ESO\,243-49. 

\subsection{Mass and age of the host star cluster}

The simplest and most natural interpretation for the colder, constant thermal component in HLX-1's optical emission (Figure \ref{fig:hst_allyears_lambda}, Section 4.1) is that of an old star cluster, host to the IMBH. We verify that this is a viable scenario by computing theoretical spectra of star clusters (represented by single-age stellar populations) with the {\sc{starburst99}} software package \citep{leitherer99,leitherer14}. For this calculation, we selected instantaneous star formation and Geneva tracks at solar metallicity, with Kroupa IMF. We examined a grid of characteristic cluster ages and masses, comparing them to the simple blackbody model fitted to the {\sc{HST}} data (Table \ref{tab:optical_ctp}. More detailed fitting of a stellar population model to the datapoints is beyond the scope of this work.

We show (Figure \ref{fig:cluster_vs_starburst99}) that a blackbody spectrum with $T_{\rm{bb}} \approx 4720$ K and bolometric luminosity $T_{\rm{bb}} \approx 2.6 \times 10^{39}$ erg s$^{-1}$ is comparable with the spectrum of a star cluster with an age of 5 Gyr with a mass between $\approx$2--3 $\times 10^{6} M_\odot$, or to that of a 10-Gyr-old cluster with a mass between $\approx$4--5 $\times 10^{6} M_\odot$. Clusters much younger than 5 Gyr are not a good match because their stellar population emission peaks at bluer colours. Globular clusters or ultracompact dwarfs in the 2--5 $\times 10^{6} \; M_\odot$ range are predicted to host IMBHs with a mass between $\approx$10$^{4}$--$10^{5} M_\odot$ (\citealt{graham20}; see also \citealt{2025PASA...42...68G}), consistent with the mass of the BH in HLX-1 inferred from X-ray spectral modelling.



\begin{figure}[t]
    \centering
    \begin{minipage}{0.47\textwidth}
        \centering
        \includegraphics[width=\linewidth]{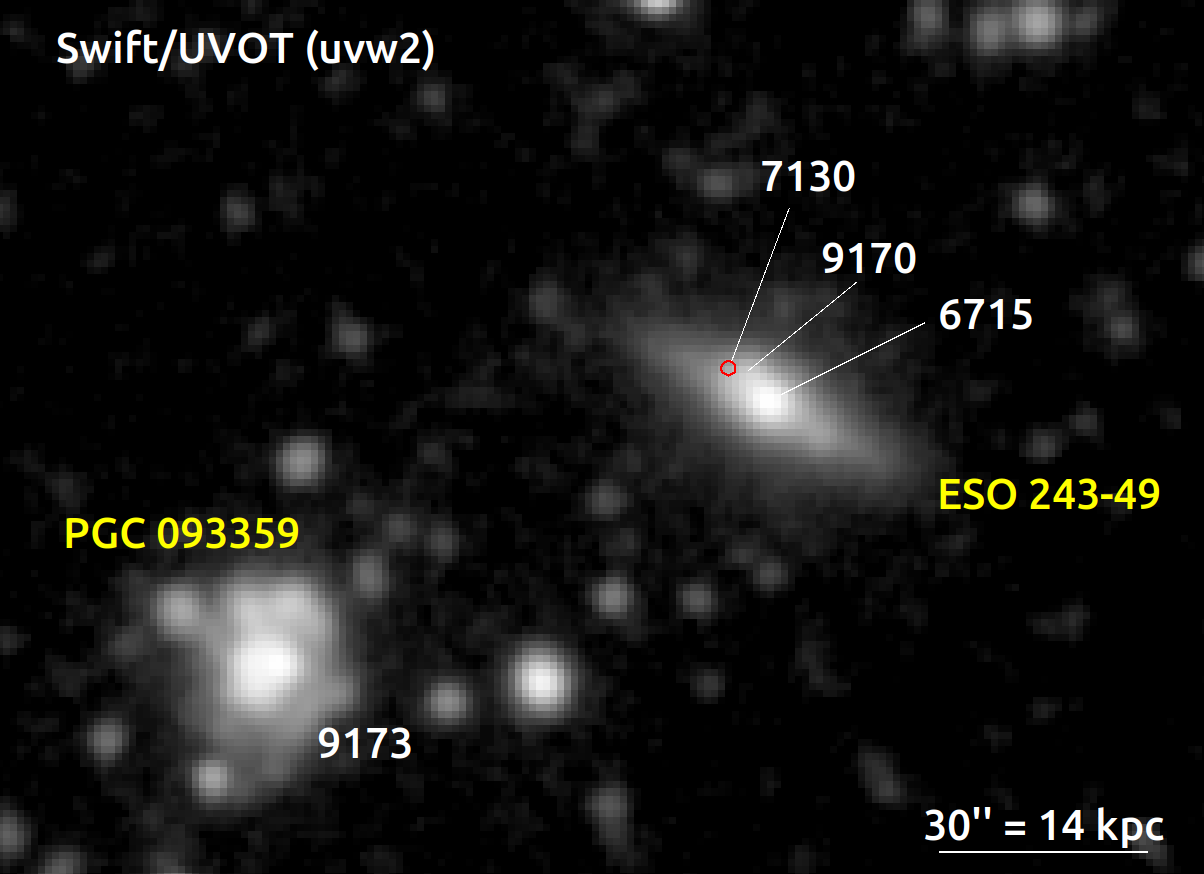}
    \end{minipage}
    \hfill
    \begin{minipage}{0.47\textwidth}
        \centering
        \includegraphics[width=\linewidth]{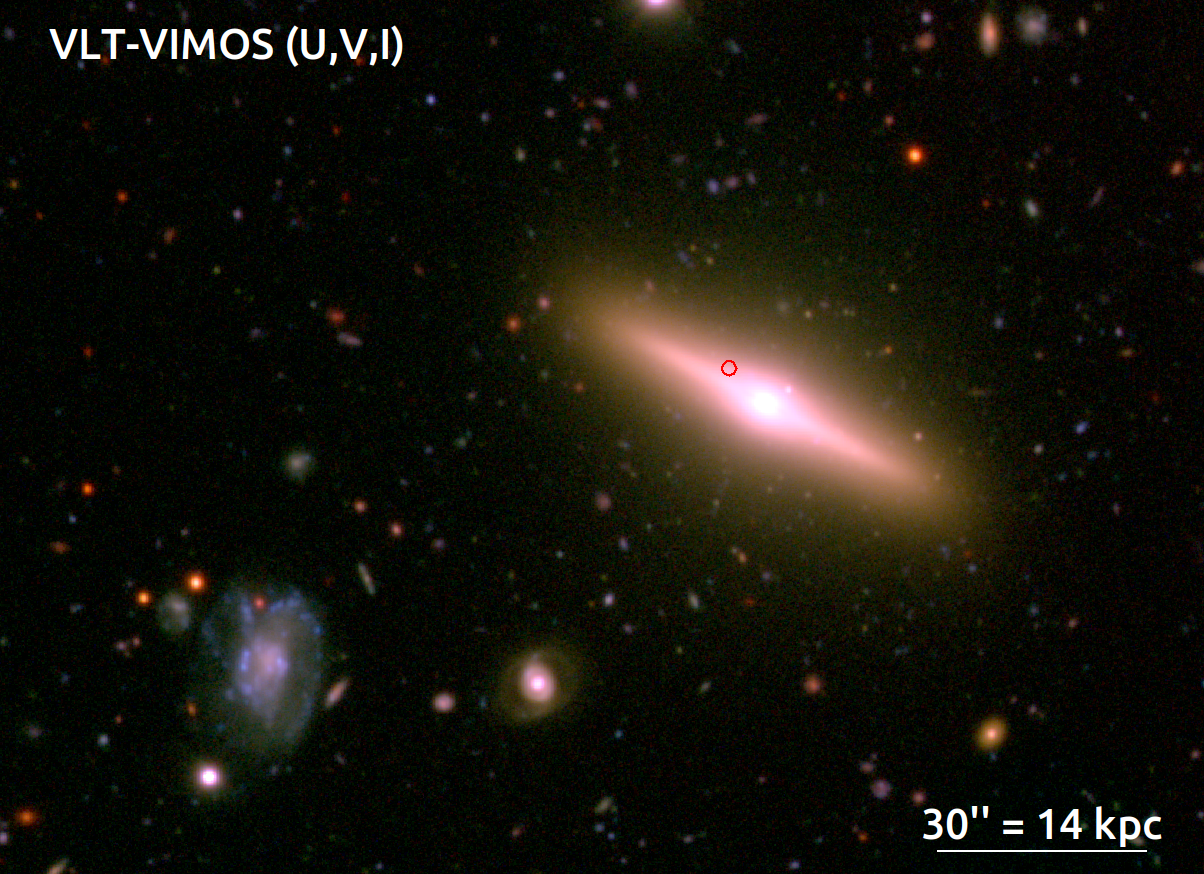}
    \end{minipage}
    \caption{Top panel: stacked {\it{Swift}}/UVOT image in the {\it{uvw2}} band (centered at $\lambda \approx 2000$ \AA), also showing excess emission between HLX-1 and the nucleus of ESO\,243-49, consistent with the flux measured from the {\it{HST}}/ACS F140LP image. Moreover, the image highlights the presence of a neighbouring late-type galaxy, PGC\,093359, $\approx$80$^{\prime\prime}$ south-east of ESO\,243-49, with a distorted structure, active star formation, and the same redshift ($cz \approx 9170$ km s$^{-1}$: \citealt{1992AJ....104..495M}) as the far-UV region near HLX-1. The position of HLX-1 is marked by a red circle. We have labelled the heliocentric radial velocities (in km s$^{-1}$) of the four main actors of our story: ESO\,243-49, HLX-1, the far-UV-emitting region between ESO\,243-49 and HLX-1, and PGC\,093359. Bottom panel: the same field seen by the VIMOS camera on the VLT (red = $I$ band; green = $V$ band; blue = $U$ band).} 
    \label{fig:uvw2_vlt}
\end{figure}

\subsection{Origin of the far-UV emitting region}


The {\it{HST}}/ACS image in the far-UV F140LP band (Figure \ref{fig:nebula_hst}) suggests that the UV-emitting structure between HLX-1 and the nucleus of ESO\,243-49 is resolved into individual clusters as well as diffused emission around and between them, with most clusters lining up close to the outer rim. The most plausible physical interpretation is a dwarf starburst galaxy \citep{lee09}, or even a collisional ring galaxy \citep{appleton96}, with star formation recently triggered and expanding outwards as a result of a tidal interaction or collisional event. First, we will estimate its current star formation rate (SFR), then we will try to understand whether this dwarf galaxy is related to HLX-1.


\begin{figure}[t]
    \centering 
    \includegraphics[width=0.47\textwidth]{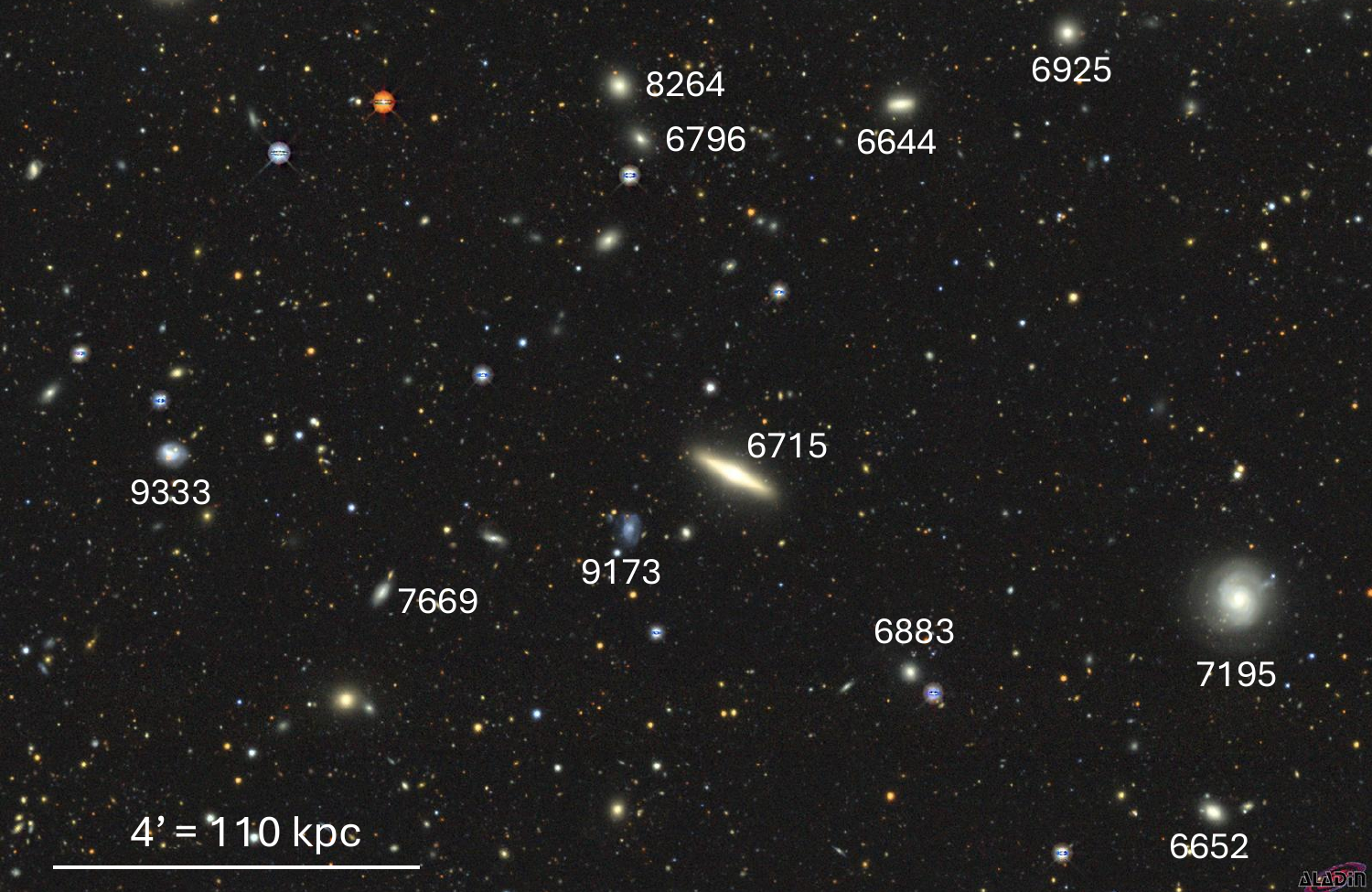}
    \caption{Larger field around ESO\,243-49 ($\approx$15$^{\prime} \times 10^{\prime}$) from the DESI Legacy Imaging Surveys Data Release 10 \citep{dey19}, with heliocentric recession speeds $cz$ (km s$^{-1}$) labelled next to some of the likely galaxy members of Abell 2877 (values from NED and \citealt{1992AJ....104..495M}).  
    }
    \label{fig:aladin_large}
\end{figure}



To estimate the SFR of the dwarf galaxy, we use again the {\sc{starburst99}} package. We selected continuous star formation, with a Large-Magellanic-Cloud abundance ($Z = 0.008$), more suitable to dwarf galaxies. We then compared the simulated spectra with the two observational datapoints from {\it{HST}}/ACS and {\it{Swift}}/UVOT, namely the extinction-corrected luminosity density $L_{\lambda} = (1.2\pm0.1) \times 10^{38} \;  \text{erg~s}^{-1} \text{\AA}^{-1}$ at $\lambda_{\rm{pivot}} \approx 1519$ \AA, and $L_{\lambda} = (0.48\pm0.02) \times 10^{38} \;  \text{erg~s}^{-1} \text{\AA}^{-1}$ at $\lambda_{\rm{pivot}} \approx 2055$ \AA. The steep spectral slope between the two wavelengths, the lack of detections in visual broadband images, the Balmer and [O\,{\footnotesize{III}}] emission lines \citep{webb17} and the small size of the star-forming region point to a young age of the starburst episode ($< 100$ Myr). For example, a starburst age of 5 Myr implies SFR $\approx$0.02 $M_\odot$ yr$^{-1}$ (an order of magnitude lower than in the Large Magellanic Cloud), while an age of 50 Myr implies SFR $\approx$0.01 $M_\odot$ yr$^{-1}$.

A more interesting question is what triggered the starburst. We exclude a direct interaction with ESO\,243-49: apart from the large velocity discrepancy (Figure \ref{fig:nebula_hst}), the latter does not show any sign of recent tidal interaction (see for example its thin, undisturbed dust lane). There is instead a larger, tidally disturbed, star-forming galaxy (PGC 093359) $\approx$50 kpc south-east of the mysterious dwarf, with the same projected recession speed (Figure \ref{fig:uvw2_vlt}). 
The star formation rate of PGC 093359 was estimated from the observed {\it{GALEX}} FUV and {\it{WISE}} W4 (22 $\mu$m) luminosities using the hybrid calibration of \cite{hao11}, as summarized by \cite{kennicutt12}. Adopting a distance of 137 Mpc from its recession speed, we derive SFR $= (0.35 \pm 0.15) M_\odot$ yr$^{-1}$; the quoted uncertainty is dominated by the intrinsic scatter of the calibration.
PGC 093359 and the dwarf projected near HLX-1 could be physically interacting, well behind ESO\,243-49. Only deep 21-cm observations (not yet available) would provide a clear answer, if they show a connecting structure of H\,{\footnotesize{I}} gas.


There is an alternative scenario that, although apparently unlikely, we cannot yet rule out: HLX-1 and its host star cluster could be related to the starburst dwarf. We need to consider the possibility that the presence of a rare X-ray-active IMBH in front of a rare starburst dwarf may not be entirely a coincidence. Abell\,2877 contains galaxies with a broad distribution of recession speeds (Figure \ref{fig:aladin_large}, and \citealt{1992AJ....104..495M}), from $\approx6700$ km s$^{-1}$ to $\approx9300$ km s$^{-1}$, projected in the same region of sky. It is not known whether such velocity spread is due to a projection effect of distinct and non-interacting clusters with different cosmological redshifts, or instead Abell\,2877 is composed of co-spatial, intersecting substructures with different proper motion (for example groups of galaxies infalling towards the cluster core from opposite directions). 
Moreover, we have already mentioned that the only velocity measurements of HLX-1 rely on a single emission line during two X-ray outbursts, a line possibly emitted from an outflow, and blueshifted. If so, the true systemic velocity of HLX-1 might be closer to the speed of the nearby starburst dwarf and of PGC 093359 than to that of ESO\,243-49. In this speculative scenario, HLX-1 and its host star cluster (perhaps an ultracompact dwarf) may have been the bullet that interacted with a gas cloud or gas-rich dwarf, and triggered the current burst of star formation, creating a collisional ring dwarf galaxy. A nearby example of such collisional systems is Kathryn's Wheel \citep{parker15,paliya24}, with its ring of star-forming knots resembling those seen near HLX-1. An alternative, equally speculative version of this scenario is that HLX-1, its apparently adjacent starburst dwarf, and PGC 093359 are physically related and were involved in a collision/ejection process.

\section{Conclusions}

\begin{itemize}[noitemsep, topsep=0pt, parsep=0pt, partopsep=0pt, leftmargin=8pt]
\item We resolved the point-like X-ray emission of HLX-1 in the low state (post-2017) from the extended emission of ESO\,243-49, and quantified their relative contributions. We found that the ongoing X-ray detections in {\it{Swift}}/XRT are at least 90\% dominated by the galaxy emission. 
\item We modelled the evolution of the UV/optical/IR emission of HLX-1 over multiple years. We found that the SED can be decomposed into a decreasing, bluer blackbody component plus a constant, redder blackbody component.
\item We proposed that the red component is consistent with the stellar emission of a star cluster with an age of a few Gyr and a mass of a few $\times 10^6 M_\odot$. This is indeed the type of globular cluster expected to host a $\sim$10$^4 M_\odot$ BH, consistent with the standard interpretation of HLX-1. Even if HLX-1 resides in an old compact stellar environment, it does not rule out the possibility of more recent dynamical interactions, such as a satellite galaxy accretion or stripping.  
\item Post-2017 observations show that the X-ray luminosity has fallen below the blue/UV component of the optical luminosity, which has instead reached a plateau, analogous to those seen in the late-time evolution of TDEs. This is inconsistent with a previously proposed scenario in which such emission comes from direct reprocessing of the X-ray irradiation. The decreasing blackbody radius associated with a slight increase in temperature point to the shrinking photosphere of an outflow, a model similar to those proposed for the UV emission in thermal TDEs at late times (hundreds of days after the event).
\item In view of the similarity between the late-time evolution of the thermal UV emission in HLX-1 and several TDEs, we re-examined the X-ray flaring behaviour of HLX-1 at earlier times. We proposed that the oscillatory behaviour predicted by TDE models (caused by radiation pressure instabilities) provides a compelling explanation for the observed recurrence timescales and luminosity behaviour of HLX-1 before 2017. We suggested that the observed outbursts are a characteristic instability during the decline of the fall-back rate, and are not the signature of a partial TDE from a star on a very eccentric orbit.
\item In this scenario, HLX-1 enabled us to follow the evolution of an IMBH TDE from an X-ray-bright, disk-dominated earlier phase (pre-2017), to the current cooling-envelope phase where the accretion rate has dropped below the threshold required to sustain radiatively efficient disk accretion.
\item We summarised the evidence for recent star formation in a nearby dwarf galaxy, possibly triggered by tidal interactions, with an SFR of $\sim$0.01--0.02 $M_\odot$ yr$^{-1}$. An association of HLX-1 with this star-forming dwarf remains unlikely (give the apparent velocity discrepancy) but not entirely impossible, and deserves further investigation.
\item To make further progress, an optical spectrum of HLX-1 in the current low state, from {\it{JWST}} or from a 10-m class ground-based telescope, is needed to verify the systemic velocities of the expanding, hot gas (emission lines) and host star cluster (absorption lines). Deep H\,{\footnotesize{I}} mapping of the surrounding field is needed to find possible tidal bridges or connecting features between HLX-1, ESO 243-49, the star-forming dwarf or the larger nearby, tidally distorted star-forming galaxy PGC 093359.

\end{itemize}
In conclusion, thanks to its multi-decade, multi-band coverage, HLX-1 can now be understood as a unique laboratory for understanding IMBHs, TDE evolution, and galaxy interaction processes.

\vspace{1cm}

This research has made use of data and software provided by the High Energy Astrophysics Science Archive Research Center (HEASARC), which is a service of the Astrophysics Science Division at NASA/GSFC.
This research is based on observations made with the NASA/ESA Hubble Space Telescope obtained from the Space Telescope Science Institute, which is operated by the Association of Universities for Research in Astronomy, Inc., under NASA contract NAS 5-26555. 
This work made use of data supplied by the UK Swift Science Data Centre at the University of Leicester.
We thank Chichuan Jin, Michela Mapelli, Christian Motch, Manfred Pakull, Beverly Smith, Doug Swartz for insightful comments and discussions on HLX-1 over the years.
RS was supported by the INAF grant 1.05.23.04.04.
RS also thanks the School of Physics at the University of Sydney, and the Institute of High Energy Physics (Beijing), for hospitality during part of this work.



\bibliography{paper}{}
\bibliographystyle{aasjournalv7}


\end{document}